\documentclass[useAMS, usenatbib]{mnras}
\usepackage{graphicx}
\usepackage{amssymb}
\usepackage{booktabs}
\usepackage{amsmath}
\newcommand\ApJ{ApJ}
\newcommand\MNRAS{MNRAS}

\newcommand\PASJ{PASJ}

\newcommand\AnA{A\&A}

\def\alf{Alfv\'en\,}
\def\bq{\begin{equation}}
\def\eq{\end{equation}}
\def\ee #1 {\times 10^{#1}}
\def\ut #1 #2 { \, \rmn{#1}^{#2}}
\def\u #1 { \, \rmn{#1}}

\def\persec {\, \hbox{s}^{-1}}

\let\grad=\nabla
\newcommand\cross{\bmath{\times}}

\newcommand\bm{\bmath}
\def\curl{{\grad \cross}}
\def\div #1 {\grad \cdot #1}

\def\bb{\bm{b}}

\def\v{\bmath{v}}

\def\v{\bmath{v}}

\def\vi{\bmath{v}_i}
\def\vj{\bmath{v}_j}
\def\ve{\bmath{v}_e}
\def\vi{\bmath{v}_i}

\def\vn{\bmath{v}_n}

\def\B{\bmath{B}}

\def\va{v_A}

\def\E{\bmath{E}}            % E
\def\vx{v_{x}}
\def\vy{v_{y}}
\def\vz{v_{z}}

\def\rl{{R_L}^*}
\def\J{\bmath{J}}

\def\sigv{<\sigma v>}

\def\kon{K\"onigl }
\newcommand\logten{\ensuremath{\log_{10}}}
\newcommand{\delt} [1] {\frac{\partial #1}{\partial t}}

\newcommand{\tenq}[1]{\hbox{\oalign{$\bm{#1}$\crcr\hidewidth$\scriptscriptstyle\bm{\approx}$\hidewidth}}}

\def\om{{\omega}}
\def\omA{{\omega_A}}
\def\omW{\omega_W}
\def\omH{\omega_H}
\def\omP{\omega_P}
\def\omgh{\omega_{gh}}
\def\omgw{\omega_{gw}}
\def\oma2{\omega_2}
\def\om4{\omega_4}
\def\bom{\bar{\omega}}
\def\bom2{\bom^2}
\title{The non-ideal finite Larmor radius effect in the solar atmosphere}
\author[B.P.Pandey and Mark Wardle]
        {B.P. Pandey and Mark Wardle \\
{School of Mathematical and Physical Sciences, Macquarie University, Sydney, NSW 2109, Australia} }
\date{\today}
\pagerange{\pageref{firstpage}--\pageref{lastpage}}
\pubyear{2021}
\begin{document}
\maketitle
\label{firstpage}
\begin{abstract}
The dynamics of the partially ionized solar atmosphere is controlled by the frequent collision and charge exchange between the predominant neutral Hydrogen atoms and charged ions. At signal frequencies below or of the order of either of the collision or charge exchange frequencies the magnetic stress is {\it felt} by both the charged and neutral particles simultaneously. The resulting neutral-mass loading of the ions leads to the rescaling of the effective ion-cyclotron frequency-it becomes the Hall frequency, and the resultant effective Larmor radius becomes of the order of few kms. Thus the finite Larmor radius (FLR) effect which manifests as the ion and neutral pressure stress tensors operates over macroscopic scales. Whereas parallel and perpendicular (with respect to the magnetic field) viscous momentum transport competes with the Ohm and Hall diffusion of the magnetic field in the photosphere-chromosphre, the gyroviscous effect becomes important only in the transition region between the chromosphere and corona, where it competes with the ambipolar diffusion. The wave propagation in the gyroviscous effect dominated medium depends on the plasma $\beta$ (a ratio of the thermal and magnetic energies). The abundance of free energy makes gyro waves unstable with the onset condition exactly opposite of the Hall instability. However, the maximum growth rate is identical to the Hall instability. For a flow gradient $\sim 0.1 \,\mbox{s}^{-1}$ the instability growth time is one minute.  Thus, the transition region may become subject to this fast growing, gyroviscous instability.
\end{abstract}


\section{Introduction}
Except for the corona, the solar atmosphere is partially ionized with varying degree of ionization in the photosphere, chromosphere and transition region between the chromosphere and corona. For example, the photosphere is weakly ionized whereas the upper transition layer (closer to the coronal boundary) is highly ionized with the partially ionized chromosphere sandwiched in-between. The plasma particles undergo frequent collision and charge exchange with the sea of predominantly neutral hydrogen atom across the magnetically threaded, stratified layers of this partially ionized gas. Thus the diffusion of the magnetic field in the solar atmosphere is non-ideal and the Ohm, Hall and ambipolar diffusion operates at the various level of this stratified plasma. For example in the photosphere where the plasma is weakly ionized and weakly magnetized, frequent collision with the neutral stops the plasma particles from drifting with the magnetic field. Thus, depending on the field strength, Ohm may dominate all other diffusion in the photosphere. With increasing altitude and thus with increasing fractional ionization and ion-magnetization (measured by the ion-Hall $\beta_i$-a ratio of the ion-cyclotron and ion-neutral collision frequencies; defined below) other non-ideal MHD effect kicks in the photosphere-chromosphere region \citep{PW06, PW08, P08}. 

Only ambipolar diffusion seems to matter with increasing ion-magnetization in the chromosphere and transition ($\sim 10^2-10^3\,$km) region. Since this region is a gateway to the mass and energy transport to the corona, it is not surprising that the role of ambipolar diffusion in dissipating the wave energy towards generating non--thermal source of heating has been the focus of recent research  \citep{D11, M11, Z11a, Z11b, Z12, Z13, KC12, L14, G14, S09, S15, CK15, S16, K17, M17, MS17, B18, CK18, RC19, M21, K21, RC21}. Thus the canonical picture that emerges from the recent research suggests that with increasing altitude the solar atmosphere changes from weakly ionized and weakly magnetized Ohm dominated photosphere to moderately ionized and highly magnetized ambipolar dominated chromosphere with the overlapping Ohm--Hall and Hall--ambipolar regions sandwiched in the middle \citep{PW12, PW13}. Even in the transition region between the chromosphere and corona where the plasma number density exceeds the neutral density by orders of magnitude \citep{F93} ambipolar diffusion remains dominant non--ideal MHD effect albeit with  decreased strength compared to the chromosphere. However, as we shall see below, this picture is not entirely accurate. 

Both the neutral and plasma particle participate in the momentum transport due to parallel, perpendicular and gyro FLR viscosity.  While in the photosphere-chromosphere region, parallel and perpendicular viscosity due to neutral may become important, in the transition region, when the plasma number density exceeds the neutral number density such that the ratio of the two densities are of the order of ion--magnetization (the ion--Hall parameter), which is given by the ratio between the ion--cyclotron $\omega_{ci}$, 
\bq
\omega_{ci}=\frac{e\,B}{m_i\,c}\,,
\label{eq:cycf}
\eq
(where $ e\,,B\,,m_i\,,c$  represents the charge, magnetic field, ion mass and speed of light respectively) and ion--neutral ($\nu_{in}$) collision frequencies, i.e., 
\bq
\beta_i=\left(\frac{\omega_{ci}}{\nu_{in}}\right)\,,
\label{eq:IhB}
\eq  
another non--ideal MHD effect, namely the gyroviscous effect, due mainly to ions, is lurking on the horizon.\footnote{The ion-neutral collision frequency, $\nu_{in}$ can be easily generalized to include the ion-ion,  and ion-electron collision frequencies (Appendix A) in the $\beta_i$ definition.} Therefore, with increasing ionization, the finite Larmor radius (FLR) cross field effect, which owing to the presence of unmagnetized ($\beta_i<1$) ions appears as Hall effect in the upper photosphere, and lower chromosphere, is reborn again in the transition region except this time it appears as gyroviscous effect due to ion magnetization ($\beta_i>1$). In table 1, we summarize the list of frequently used symbols in this paper.

The solar photosphere--chromosphere and transition region is a partially ionized mixture of ions and neutrals with the ionization fraction
\bq
X_e = \frac{n_e}{n_n}\,,
\eq  
as low as $10^{-4}$ around the photospheric temperature minimum and as high as $\gtrsim 10^{4}$ in the upper transition region \citep{F93}. Here $n_e$ and $n_n$ are the number densities of the electrons and neutrals respectively. 

Owing to the high collisionality, ions (number density $n_i$) could be tightly coupled to the neutrals so that they pick up the neutral inertia, gaining an effective mass
\bq
m_i^{*} = \frac{m_i\,n_i + m_n\,n_n}{n_i}\,,
\label{eq:dms}
\eq
in the process and thus are unable to fully respond to the changes with frequencies in excess of Hall frequency \citep{PW06, PW08}
\bq
\omH = \left(\frac{m_i}{m_i^{*}}\right)\,\omega_{ci}\,.
\label{eq:hallf}
\eq
Implied in this description is the assumption that the collisions are able to provide strong coupling between the ions and  neutrals, i.e.,
\bq
\omega \lesssim \left(\frac{\rho}{\rho_n}\right)\,\nu_{ni} \,.
    \label{eq:omegax}
\eq 
Here $\omega$ is the signal frequency, $\rho = \rho_i + \rho_n$ is the bulk mass density and  $\rho_{i\,, n} = m_{i\,,n}\,n_{i\,,n}$ are the ion and neutral mass densities.

It becomes immediately clear from Eq.~(\ref{eq:dms}) that the dressed ion mass increases dramatically in a weakly ionized environment since $ m_i^{*} \approx m_n/X_e$ (here we have assumed that the plasma is  quasineutral i.e., $n_i \approx n_e$). As a result the Larmor radius in a weakly ionized medium is a function of the fractional ionization \citep{P13} 
\bq
R_L^{*}=\left(\frac{c_s}{\omH}\right) \sim X_e^{-1}\,R_L\,, 
\label{eq:Lar}
\eq
and will be much larger than its counterpart,  $R_L=c_s/\omega_{ci}$ in the fully ionized medium. Here $c_{s}=\sqrt{k_B\,T/m_i}$ is the ion thermal/sound speed and $k_B\,,T$  is the Boltzmann constant and temperature respectively. With increasing ionization  $R_L^{*}$ approaches $R_L$.  

The ratio of the effective Larmor radius $R_L^{*}$ and Hall scale $L_H = \va/\omH$ becomes
\bq
\left(\frac{R_L^{*}}{L_H}\right) \sim \left(\frac{c_{s}}{\va}\right)
\sim \sqrt{\beta}\,.
\label{eq:hBeta1}
\eq 
Here $\va = B/\sqrt{4\,\pi\,\rho}$ is the \alf speed in the bulk fluid and plasma beta 
\bq
\beta=\frac{2\,c_s^2}{\va^2}\,,
\eq
is the ratio of the plasma thermal and magnetic energies. 

Adopting an altitude dependent profile of the magnetic field
\bq
B = B_0\,\left(\frac{n_n}{n_0}\right)^{0.3}\,,
\label{eq:scl}
\eq 
that captures the essential height variation of the observed fields in the flux tubes \citep{M97} and taking the neutral number density and fractional ionization from Model C \citep{VAL81} shows that the Ohm, Hall and ambipolar diffusion dominates various altitude of the atmosphere \citep{PW13}. The improved model of \cite{F93} also gives similar diffusion coefficient in the photosphere--chromosphere and transition region (Appendix C).

\begin{figure}
\includegraphics[scale=0.35]{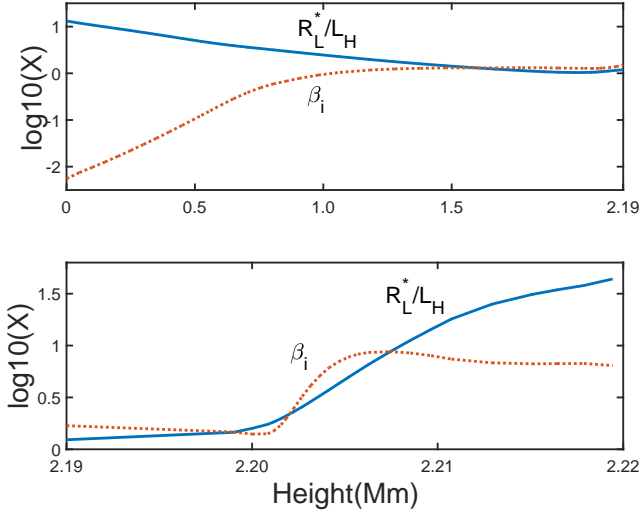}
\caption{The ratio $R_L^*/L_H$  and ion--Hall parameter $\beta_i$ are plotted against height for $B_0=100\,\mbox{G}$. The altitude dependence of the magnetic field is taken from Eq.~(\ref{eq:scl}) and data from table 2 (model C) of \citep{F93} have been used for density and temperature. For better resolution, the upper transition region has been shown separately in the lower panel.}
 \label{fig:FN1}  
\end{figure}
\begin{figure}
\includegraphics[scale=0.35]{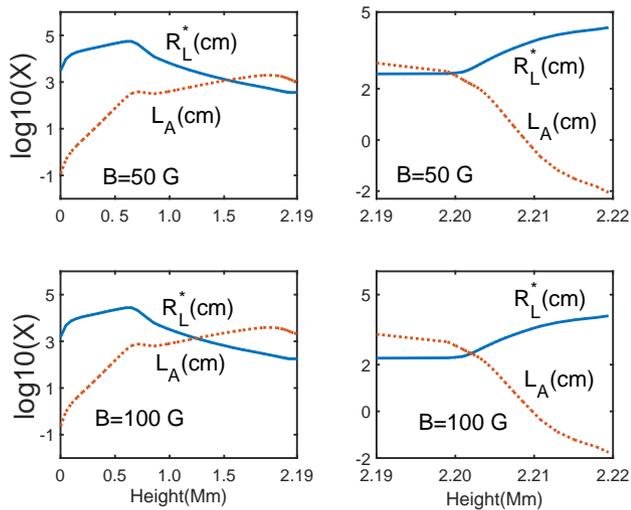}
\caption{The Larmor radius $\rl(cm)$  and ambipolar length scale $L_A(cm)$ are plotted for $B_0=50\,\mbox{G}$ (top panels), and $B_0=100\,\mbox{G}$ (bottom panels) fields. For better resolution, the upper $32\,\mbox{km}$ of the transition region (right hand panel) has been plotted separately. The altitude dependence of the magnetic field  and the density and temperature profiles are same as in the previous figure.}
 \label{fig:FN2}  
\end{figure}

In Fig.~(\ref{fig:FN1}), we plot the ratio $R_L^*/L_H$ using model C of \cite{F93} and assuming $B_0=100\,\mbox{G}$ at the footpoint. 
For Hall $\beta_i$, we have used $\nu_i$ [Eq.~(\ref{eq:cfx}), Appendix A] which apart from $\nu_{in}$ also includes $\nu_{ii}$ and $\nu_{ie}$. It is clear that $R_L^*$ dominate $L_H$ in the photosphere, lower chromosphere and then in the transition region. 
Thus, as $\beta_i<1$ in the photosphere-lower chromosphere, or $\beta_i>1$ in the middle and upper chromosphere, the magnetic diffusion (due to Hall, or ambipolar) will be accompanied by the parallel and perpendicular viscous momentum transport. However, in the transition region, where ions are magnetized ($\beta_i>1$ ), $R_L^*/L_H \gg 1 $, and as we shall see from Fig.~(\ref{fig:FN2}), $R_L^*/L_A \gg 1 $ (ambipolar scale, $L_A=\va/\nu_{ni}$), ambipolar diffusion disappears and gyroviscous momentum transport becomes important.   

Kinetic theory provides an appropriate framework to investigate the role of FLR effects at any wavelength, $k^{-1}$ (here $k$ is the wavenumber) on the plasma dynamics \citep{RKR62}. However, in the various limiting cases, i.e., for a given wavenumber $k$ when $k\,R_L>1$, or $k\,R_L \lesssim 1$, magnetohydrodynamic (MHD) framework is easily extendable to include the FLR effect by either modifying the induction \citep{HH88}, or momentum equation \citep{RT62}. When $k\,R_L>1$ only FLR modification to the MHD of fully ionized plasma is the appearance of Hall term in the induction equation. However, as we shall see, in a partially ionized plasma the parallel and perpendicular  viscosity due to neutrals becomes as important.  When $k\,R_L<1$  perpendicular and  gyro viscosity appears in the extended MHD framework \citep{BR65}.   

Although in partially ionized plasmas, the measure of ion magnetization is given by the ion--Hall parameter $\beta_i$, it is not difficult to see that the large larmor radius limit, $k\, R_L^{*}>1$ and  $\beta_i<1$ is equivalent. In order to see this, lets equate the neutral--ion collision mean free path, 
\bq
\lambda_{\mbox{mfp}} \sim \frac{c_s}{\nu_{ni}}\,,
\eq
with the fluctuation wavelength, $k^{-1}$. This yields 
\bq
k\,R_L^{*} \sim \frac{1}{\beta_i}\,.
\eq
Therefore, $k\, R_L^{*}>1$, and $\beta_i<1$ are equivalent. Clearly, the reduced single fluid description of the partially ionized plasma by \cite{PW06, PW08} anticipates large larmor radius ($k\,R_L^{*} > 1$, or $\beta_i<1$) correction to the induction equation although parallel and perpendicular viscosity (which is mainly due to neutrals) was not considered. Further the small larmor radius, $k\,R_L^{*}<1$, (or $\beta_i>1$) limit which is dominated by the perpendicular and gyroviscous stress  has not been considered. As we shall see this correction becomes quite important in the upper chromosphere, and transition region between the chromosphere and corona.       

To summarize, in a partially ionized medium, due to their frequent collision with the neutrals, ions acquire an effective mass $m_i^*$. As a result of this {\it mass loading by the neutrals} the ion--cyclotron frequency,  $\omega_{ci}$ becomes Hall frequency $\omH$ \citep{PW06, PW08}. Therefore, the Larmor radius of the dressed ion becomes a function of the fractional ionization of the medium. As the magnetic field cannot directly couple to the neutrals, the mass loading of the ion implies that only for the frequencies of interest that satisfies Eq.~(\ref{eq:omegax}), i.e., for the low frequency long wavelength fluctuations in the medium this physical description is valid. 

When charge exchange is either as frequent as neutral--ion collision, or is dominant, we can as well derive the above expression for the Hall frequency, Eq.~(\ref{eq:hallf}) and the modified ion Larmor radius, Eq.~(\ref{eq:Lar}) by arguing that the ion acquires an effective {\it reduced} charge rather than an effective {\it increased} mass. Recall that when the neutral Hydrogen atom loses an electron to a nearby ion (usually neutralizing it) charge exchange/transport occurs. Post exchange neutral Hydrogen becomes {\it new} ion and ion becomes {\it new} neutral Hydrogen atom. Thus, the charge exchange represents a coupling process between the ions and neutrals \citep{HW98}. In a thermalized plasma environment when there is no energy transfer between the ions and neutrals, if this exchange of identity is sufficiently rapid, the charge on the particle will flip--flop between $0$ and $q$ acquiring in the process an effective charge \citep{V16}
\bq
q_{\mbox{ef}} = \left(\frac{\rho_i}{\rho}\right)\,q\,.
\eq   
Making use of $ q_{\mbox{ef}}$ instead of $e=q$ in Eq.~(\ref{eq:cycf}) we get Eq.~(\ref{eq:hallf}), i.e. ion gyrates at Hall frequency $\omH$ around the magnetic field. Also the effective Larmor radius is now $R_L^{*}$ and not $R_L$. Only requirement is that the exchange of charge between the ions and the neutrals should be rapid enough over the dynamical time scale i.e. must satisfy Eq.~(\ref{eq:omegax}) where the collision frequency on the right hand side now will also include charge exchange frequency. 

Is the process of charge exchange rapid enough in the solar atmosphere in comparison with the dynamical timescale, $\omega^{-1}$? As the charge exchange cross-section is of the same order as the momentum exchange cross--section [Fig.~2, \cite{V16}], the charge exchange frequency is comparable to the ion--neutral collision frequency. One--dimensional, semi--empirical, non--LTE, hydrostatic model of transition region \citep{F93} assumes that the main collisional interaction between protons and hydrogen atoms is elastic charge exchange. The inequality Eq.~(\ref{eq:omegax}) is easily satisfied for the charge exchange process. Therefore, we may assume that the ions either have same charge but an effective mass, or same mass but an effective charge. These complementary viewpoint results in the identical spatial and temporal scales. Clearly the single fluid description of partially ionized solar plasma \citep{PW06, PW08} anticipates charge exchange. 
\begin{table*}
\hspace*{-4cm} 
\begin{minipage}{110mm}
\caption{List of frequently used symbols.}
\label{mathmode}
\begin{tabular}{@{}llll}
Sybmol & Explanation & Symbol & Explanation\\
\hline
$X_e$  & Ionization fraction       & $m_i\,,m_i^*$& ion and effective mass \\[2pt]
$\omega_{ci}$             & ion-cyclotron frequency        &$\omH=\frac{m_i}{m_i^*}\omega_{ci}$   & Hall frequency   \\[2pt]
$\nu_i$  &  Sum of $\nu_{ii}\,,\nu_{in}\,,\nu_{ie}$&$\nu_n$ & Sum of $\nu_{nn}\,,\nu_{ni}\,,\nu_{ne}$ \\[2pt] $c_s=\sqrt{\frac{k_B\,T}{m_i}}$             & sound speed        &$\va=\frac{B}{\sqrt{4\,\pi\,\rho}}$   & \alf speed   \\[2pt]
$R_L^{*}=\frac{c_s}{\omH}$                   & Larmor radius        &   $\beta=\sqrt{\frac{2\,c_s^2}{\va^2}}$    & plasma beta \\[2pt]
$\beta_i=\frac{\omega_{ci}}{\nu_i}$ &Ion hall beta&$\eta_{i\,,n\,0\,,1\,,2\,,3\,,4}$&viscosity coefficients\\[2pt]
$\nu_0=(\eta_{i\,0}+\eta_{n\,0})/\rho $&parallel viscosity&$\nu_{1\,,2}=(\eta_{i\,1\,,2}+\eta_{n\,1\,,2})/\rho$&perpendicular viscosity \\[2pt]
$\nu_3\,,\nu_4=(\eta_{i\,,3\,,4}+\eta_{n\,,3\,,4})/\rho$    &gyro viscosity   &        $\tenq{W}$    & strain tensor \\[2pt]
 $\tenq{W}_0$     &  parallel strain tensor&$\tenq{W}_1\,,\tenq{W}_2$ &perpendicular strain tensors\\[2pt]
$\tenq{W}_3\,,\tenq{W}_4$  & cross strain tensors&$L_A=\va/\nu_{ni}\,,L_H=\va/\omega_H$ & Ambipolar and Hall scale\\
\hline
\end{tabular}
\end{minipage}
\end{table*}

Most solar heating, with the possible exception of flares, takes place in the partially ionized chromosphere and transition region. The lower chromosphere is threaded by the strong ($\sim$ kilogauss) vertical flux tubes located in the network regions where they are observed as bright points \citep{H09}. These flux tubes starts near the foot point in the photosphere, expand with increasing height before filling the entire atmosphere and forming a canopy in the chromosphere. The transition region is threaded by a mixture of hot and cold magnetic loops \citep{AD91, D93}. As these regions are threaded by flux tubes the magnetic field plays an important role in transporting the convective energy to heat the plasma locally. It is believed that the dissipation of the \alf waves in the chromosphere and transition region is a plausible non--thermal energy source that may heat the entire corona \citep{KC12, MS17, S21}. 

As ions are highly magnetized ($\beta_i\gg1$) in the upper chromosphere and transition region, it would appear that the only non--ideal MHD effect that needs to be considered in this region is ambipolar diffusion. However, as we see from 
Fig.~(\ref{fig:FN2}), gyroviscocity may become dominant transport mechanism in the transition region. Assuming $B_0=50\,\mbox{G}$ Fig.~(\ref{fig:FN2}, top panels), or $100\,\mbox{G}$ Fig.~(\ref{fig:FN2}, lower panels), we see that the gyrovisocus scale $\rl$ dominates ambipolar scale $L_A$ up until the middle chromosphere. From the middle to the upper chromosphere $L_A$ dominates $\rl$, although for a weaker ($B_0<50\,\mbox{G}$) field, $\rl$ may dominate $L_A$ in the upper chromosphere. Finally in the transition region ($\gtrsim 2.2\,\mbox{Mm}$), neutral hydrogen density plummets, ambipolar diffusion almost disappears and $\rl/L_A\gg1$. Clearly, gyroviscous momentum transport is the dominant non--ideal MHD effect in the transition region between the chromosphere and corona. Therefore, FLR effect manifests twice in the solar atmosphere: (a) first as the parallel and perpendicular viscosity (mainly due to neutrals) together with the magnetic  (Hall) diffusion at short ($k\,R_L^{*} \sim 1/\beta_i > 1$) wavelengths in the photosphere and chromosphere, and (b) as the gyroviscosity (mainly due to ions) in the transition region when ($k\,R_L^{*} < 1$). As the gyroviscous effect causes polarized waves that propagate at scaled (by plasma $\beta$) whistler and Hall frequencies, these waves may propagate through the transition region with little, or no damping.  

Rotating features are ubiquitous in the solar atmosphere. Photospheric vortical motions often force co--localized magnetic flux tubes in the intergranular lanes (which threads the upper atmosphere) to rotate. This in turn produces observable corotating structures in the chromosphere and corona as {\it chromospheric swirls} and {\it magnetic tornadoes} respectively. These rotating magnetic structures occur over a large range of spatial scales and extend from the upper convection zone to the transition region and lower corona \citep{KW17}. A small (granular) scale vortices in the quiet Sun has also been observed recently \citep{P16}. The small scale ($<0.4-0.5\,\mbox{Mm})$) vortex motion in the moderately magnetized (network) regions are typically associated with the intergranular convective downdrafts \citep{B08, MS11}. Most vortices are small ($\lesssim 0.5\,\mbox{Mm}$) with average size of few hundred km and typical lifetime of few minutes, although large vortices $\sim 20\,\mbox{Mm}$ with lifetime $\gtrsim 20\,\mbox{min}$ have also been observed \citep{At09}. The bright points associated with the vortex motion in the intergranular lane moves with typical speed $\lesssim 2\,\mbox{km} / \mbox{s}$ \citep{W09}. All in all the vortex-—like, or swirling flows are ubiquitous in the photosphere--chromosphere \citep{B08, W09, Ba10, B10}. 

The formation of small, granular--scale vortices in quiet regions of the Sun is a generic feature of the turbulent convective sub--surface layers \citep{Z93, SN98, N09, M10}. The MHD simulations suggest that small--scale eruptions in the solar atmosphere can be driven by magnetized vortex tubes, i.e., by magnetic tornadoes \citep{M11, S11, W12, K13, WS14}. The quasiperiodic ($2-5\,\mbox{min}$) plasma eruptions in swirling tubes are generated by the turbulent convection in subsurface layers \citep{K13}. However, not all photospheric vortex flows have a chromospheric counterpart. In fact the occurrence rate of chromospheric vortex flows is much smaller (by an order of magnitude) than the corresponding rate ($3.1\times 10^3\,\mbox{Mm}^{-2}\,\mbox{s}^{-1}$) for photospheric vortices \citep{KW17}. This is because the vortex flow is also present in the chromosphere only when the footpoint of the magnetic flux tube in the photosphere coincides with the photospheric vortex tube \citep{WS14}. Therefore, the topology of the magnetic field plays crucial role in facilitating the transfer of swirling motion from the photosphere to the chromosphere.  The non-ideal MHD simulation also show that the Hall effect can generate out-of-plane velocity fields with maximum speed $\sim 0.1\,\mbox{km} / \mbox{s}$ at the interface layers between weakly magnetized light bridges and neighbouring strong field umbral regions \citep{C12}. Clearly, both the observation and numerical simulation points to the presence of shear flow at various spatial scales in the solar atmosphere. 

Heating of the corona of magnetically quiet solar atmosphere and the origin of solar wind requires an energy flux $\sim 1-3\,\times 10^3-10^4\,\mbox{erg}\,\mbox{cm}^{-2}\,\mbox{s}^{-1}$. Magnetic vortices are closely related to the local heating \citep{M11, WS14}. Solar tornadoes may facilitate the transfer of vortex energy to the chromosphere and corona and thus contribute to their heating. 

The presence of free shear flow energy in the transition region can easily make waves ustable. Therefore, we shall investigate the local stability of the waves in the transition region due to gyroviscosity by assuming that the magnetic field is immersed in the highly diffusive medium. Further, we shall compare and contrast the gyroviscous shear instability with the non--ideal Hall instability \citep{PW12, PW13}. As the solar atmosphere is highly stratified, the validity of the present local analysis is restricted to the short (with respect to the scale height, $h$) wavelength fluctuations, i.e., $k\,h \gg 1$.  As $k\sim \nu_{ni}/c_s$, and $h\sim c_s/\omega$, the requirement $k\,h \gg 1$ translates into Eq.~(\ref{eq:omegax}). Because the flux tube will be approximated by a planar geometry, our analysis is valid only for the wavelengths much smaller than the tube radius $r$, i.e., $k\,r\gg 1$.
\section{Basic set of equations and waves and instabilities in the medium}
The solar atmosphere is a partially ionized medium with neutral hydrogen atom as its main constituent in the photosphere-chromosphere. Number density of plasma particles (compared to the neutrals) remains very low in the photosphere and low and middle chromosphere. However, the plasma number density becomes comparable to the neutral number density in the upper chromosphere before exceeding it in the transition region \citep{F93}. Due to their great simplicity, a single fluid MHD like description of the multi--component solar plasma provides an optimal framework to investigate the waves and instabilities in the medium. The starting point is the following set of Braginskii equations \citep{BR65, GP04, SC09} 
\begin{equation}
\frac{\partial \rho_j}{\partial t} + \grad\cdot\left(\rho_j\,\vj\right) = 0\,,
\label{contj}
\end{equation}
\begin{equation}
\frac{d\ve}{dt}= - \frac{\nabla P_e}{\rho_e} - \frac{\nabla \cdot \Pi_e}{\rho_e}- \frac{e}{m_e}\left(\E + \frac{\ve}{c}\cross \B\right)
- \bmath{R}_{ej}\,,
\label{eeq}
\end{equation}
\begin{equation}
\frac{d\vi}{dt}=   -\frac{\nabla\,P_i}{\rho_i}-\frac{\nabla \cdot \Pi_i}{\rho_i}  + \frac{e}{m_i}\,\left(\E + \frac{\vi}{c}\cross \B\right)
- \bmath{R}_{ij}\,,
\label{ieq}
\end{equation}
\begin{equation}
\frac{d\vn}{dt}= - \frac{\nabla\,P_n}{\rho_n}-\frac{\nabla \cdot \Pi_n}{\rho_n} +
\sum_{j=e,i} \!\!\nu_{nj}\left(\vj - \vn \right)\,.
\label{neq}
\end{equation}
Here $j=i,\,e,\,n$.  We shall assume that the ions
are singly charged and adopt charge neutrality, so that $n_i \approx n_e$. The collisional terms in the above momentum equations are
\bq
\bmath{R}_{ej} = \!\! \sum_{j=i,n}\!\!\nu_{ej}\left(\ve - \vj \right)\,,\quad
\bmath{R}_{ij} = \!\! \sum_{j=e,n}\!\!\nu_{ij}\left(\vi - \vj \right)\,,
\eq
where $\nu_{ej}$ and $\nu_{ij}$ are the electron and ion collision frequencies with the $\mbox{j}^{\mbox{th}}$ particle. 

The non--diagonal viscous stress tensor is calculated from the linearized kinetic equations \citep{BR65, C86} and can be written in the component form as 
\bq
\tenq{\Pi} = \tenq{\Pi}_{\parallel} +  \tenq{\Pi}_{\perp} + \tenq{\Pi}_{\Lambda}\,.
\eq
Here $\parallel\,,\perp\,,\Lambda$ are the parallel [$\bb(\bb\cdot\nabla)$], perpendicular [$-\bf{b}\cross(\bf{b}\cross \nabla$)] and cross ($\bf{b}\cross \nabla$) terms with respect to the magnetic field direction $\bb=\B /B$. Thus the momentum transport due to Braginskii tensor $\tenq{\Pi}$  consists of (a) transport along the magnetic field lines--$\tenq{\Pi}_{\parallel}$, i.e., field--free transport, (b) perpendicular transport, $\tenq{\Pi}_{\perp}$ which is smaller (comparable)  to the field--aligned transport by $\sim 1/\beta_i^2$ for the ions (neutrals), and (c)  gyro, or cross--field transport, $\tenq{\Pi}_{\Lambda}$  which is smaller (negligible) by the factor $\sim 1/\beta_i$ for the ions (neutrals) compared to the field--aligned transport.  

In terms of rate of strain $\tenq{W}$, stress tensor $\tenq{\Pi}$ is
\bq
\tenq{\Pi} = - \eta_0\,\tenq{W}_{0} - \eta_1\,\tenq{W}_{1} - \eta_2\,\tenq{W}_{2} + \eta_3\,\tenq{W}_{3} + \eta_4\,\tenq{W}_{4}\,,
\label{eq:vstx}
\eq
The various ion and neutral $\eta$ coefficients are related to the plasma pressure, magnetization  and collsion frequencies (Appendix A). Tensor $\tenq{W}_{j}$ in the above Eq.~(\ref{eq:vstx}) is expressed in terms of $\tenq{W}$ [Eq.~\ref{eq:stn}] via Eq.~(\ref{eq:wab}) (detail in the the Appendix A). First term in the above Eq.~(\ref{eq:vstx}) is the parallel stress, second and third is the perpendicular stress and the last two terms is the cross stress which is called  gyroviscous stress.

\begin{figure}
\includegraphics[scale=0.35]{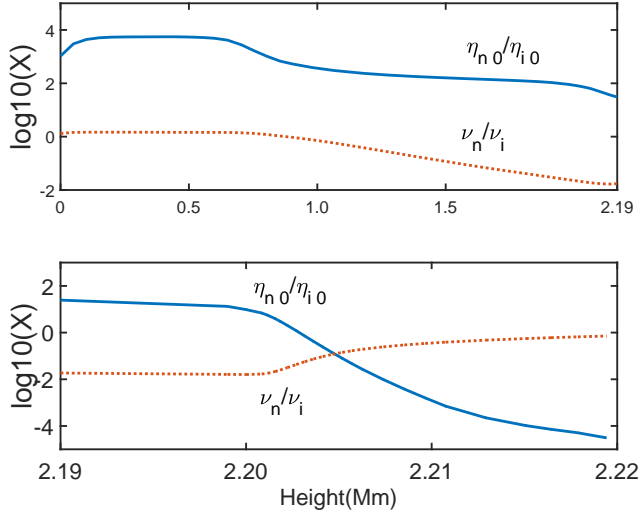}
\caption{The ratio $\eta_{n\,0}/\eta_{i,0}$ (solid curve) and $\nu_n/\nu_i$ (dotted curve) are plotted against height for   $B_0=100\,\mbox{G}$  for the same altitude dependence of the magnetic field and density and temperature as in the previous figure.}
 \label{fig:Fpar}  
\end{figure}

In order to assess the relative importance of these viscous terms, we compare them with the neutral pressure gradient term. Thus in the transition region, comparing the field-aligned viscous term to the neutral pressure gradient yields 
\bq
\frac{|\nabla \cdot \Pi_{\parallel}|}{|\nabla\,P_n|} \sim \frac{\eta_{i\,0}|\nabla\v|}{P_n} \sim \frac{P_i}{P_n}\frac{|\nabla\v|}  {\nu_i}\sim X_e\,, 
\eq
where we have assumed that $|\nabla\v|\sim \nu_i$, $\eta_{n\,0}/\eta_{i\,0}<1$  and $\nu_n/\nu_i\lesssim 1)$ (Fig.~\ref{fig:Fpar}). In the photosphere-chromosphere region however, $\eta_{n\,0}/\eta_{i\,0}>1$ and parallel viscosity transport is mainly due to neutrals. Thus
\bq
\frac{|\nabla \cdot \Pi_{\parallel}|}{|\nabla\,P_n|} \sim \frac{\eta_{n\,0}|\nabla\v|}{P_n} \sim \mathcal{O}(1)\,.
\eq

Similar ratio for the perpendicular term in the transition region is
\bq
\frac{|\nabla \cdot \Pi_{\perp}|}{|\nabla\,P_n|} \sim \frac{X_e}{\beta_i^2}\,,
\eq
while in the photosphere-chromosphere it becomes
\bq
\frac{|\nabla \cdot \Pi_{\perp}|}{|\nabla\,P_n|} \sim \mathcal{O}(1)\,.
\eq

For the gyroviscous term we have
\bq
\frac{|\nabla \cdot \Pi_{\Lambda}|}{|\nabla\,P_n|} \sim \left(\frac{P_i}{P_n}\right)\frac{1}{\beta_i}\,, 
\eq
Clearly if this ratio is $\sim \mathcal{O}(1)$ the gyroviscous term in the momentum equation will become as important as the pressure gradient terms. In a thermalized ($T_i=T_n$) plasma environment, this can be recast in terms of fractional ionization, $X_e \sim \beta_i$. 
\begin{figure}
\includegraphics[scale=0.35]{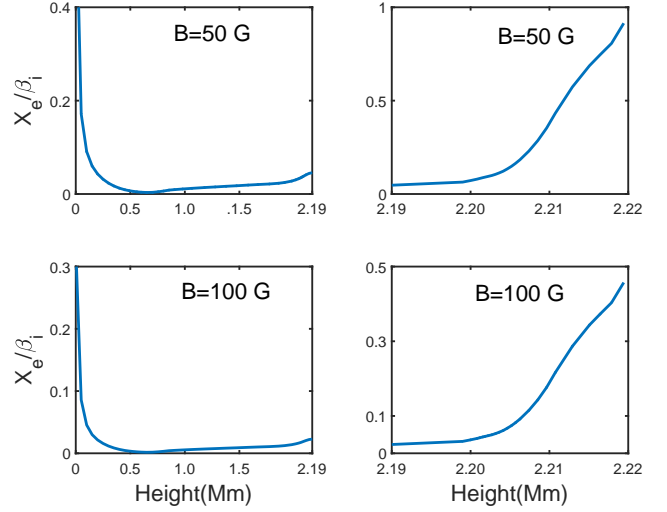}
\caption{The ratio $X_e/\beta_i$ is plotted for $B_0=50\,\mbox{G}$ (top panels) and  $B_0=100\,\mbox{G}$ (bottom panel). Other 
parameters are same as in the previous figure.}
 \label{fig:FN3}  
\end{figure}
We see from Fig.~(\ref{fig:FN3}) that for $B_0=50\,\mbox{G}$, $X_e/\beta_i \sim 1$ whereas for 
$B=100\,\mbox{G}$ field, $X_e/\beta_i \sim 0.5$. Thus the role of gyroviscous effect in the momentum transport can not be ignored in the transition region.   

The various component of Braginskii tensor $\tenq{\Pi}_i$ and $\tenq{\Pi}_n$ is given in the Appendix A.  Relative contribution of the electron viscosity to the stress tensor is important only if the electron temperature $T_e$ satisfies \citep{Z02} $T_e \sim (m_i/m_e)^{1/5}\,T$, or $T_e\gtrsim 4.49\,T$. However, as the partially ionized solar plasma is in thermal equilibrium ($T_e=T_i=T_n=T$),  electron viscosity is unimportant and thus has been neglected. We shall also neglect the electron inertia.  Defining the mass and current densities and velocity of the bulk fluid as
\begin{eqnarray}
\rho &=& \rho_e + \rho_i + \rho_n \approx \rho_i + \rho_n\,,
\nonumber\\
\J &=& e\,n_e\,\left(\vi - \ve\right)\,,
\nonumber\\
\v &=& (\rho_i\,\vi + \rho_n\,\vn)/\rho \equiv (1-D)\,\vi + D\,\vn\,,
\label{rho}\,.
\end{eqnarray}
where 
\bq
    D = \left(\frac{\rho_n}{\rho}\right)\,,
    \label{eq:D}
\eq
the above momentum equations, in the absence of Braginskii tensor can be reduced to a single fluid, MHD like momentum equation valid for arbitrary ionization provided the dynamical frequency $\omega$ satisfies Eq.~(\ref{eq:omegax}) \citep{PW06, PW08}. When ions are weakly magnetized ($\beta_i\lesssim 1$)  parallel and perpendicular (comparable to the parallel) viscosities mainly contribute to the viscous stress (Appendix A, Fig.~\ref{fig:FV}). On the other hand, when ions are highly magnetized ($\beta_i\gg1$) gyroviscosity is the main contributor to the viscous momentum transport.  

In order to incorporate $\tenq{\Pi_i}$ and $\tenq{\Pi_n}$ in the single fluid framework, we note that
\bq
\vi = \v-D\,\left(\vn-\vi\right)\,,
\label{eq:avi}
\eq
and
\bq
D\,\left(\vn-\vi\right) \sim D\,\frac{\J\cross\B}{c} \sim D\,\frac{\va^2}{L\,\nu_{ni}}\,,
\eq
for the gradient of scale length $L$. Thus for $|\v|\sim \va$  the last term in Eq.~(\ref{eq:avi}) can be neglected provided Eq.~(\ref{eq:omegax}) is satisfied. Thus assuming $\vi\approx\v$ the strain tensor $\tenq{W}$ 
\bq
\tenq{W} = \grad\v + \left(\grad\v\right)^{T} - \frac{2}{3}\,\tenq{I}\div\v\,,
\label{eq:stn}
\eq
now correspond to the strain in the bulk fluid. 

We write $\tenq{\Pi}=\tenq{\Pi}_i+\tenq{\Pi}_n$ where $\tenq{\Pi}$ in terms of rate of strain $\tenq{W}$ is 
\bq
\tenq{\Pi} = - \rho\nu_0\,\tenq{W}_{0} - \rho\nu_1\,\tenq{W}_{1} - \rho\,\nu_2\,\tenq{W}_{2} + \rho\nu_3\,\tenq{W}_{3} + \rho\,\nu_4\,\tenq{W}_{4}\,,
\label{eq:vst}
\eq
with
\bq
\nu_0=\frac{\eta_{i\,0}+\eta_{n\,0}}{\rho}\,.
\eq
Similarly for $\nu_1\,, \nu_2\,,\nu_3\,,$ and $\nu_4$. Writing $P=P_i+P_n+P_e$ and adding the momentum equations together, we arrive at the following single fluid equations 
\bq
\frac{\partial \rho}{\partial t} + \grad\cdot\left(\rho\,\v\right) = 0\,,
\label{eq:cont}
\eq
\bq
\rho\,\frac{d\v}{dt}=  - \nabla\,P - \nabla\cdot \tenq{\Pi} + \frac{\J\cross\B}{c}\,.
\label{eq:meq}
\eq

The viscous stress term also modifies the generalized Ohms law. This leads to the appearance of an additional term in the induction equation Eq.~(25) of \cite{PW08} proportional to $(D^2/\rho_i\nu_{in})(\nabla \cdot \Pi\cross\B)$ which can be neglected with respect to the advective ($\v\cross\B$) term if
\bq
\omega\lesssim\left(\frac{\va}{c_s}\right)\left(\frac{\rho}{\rho_n}\right)\,\nu_{ni}\,.
\label{eq:gnc}
\eq
When $\rho_i\gtrsim \rho_n$ the neglect of $\grad P\cross B$ terms with respect to the $\v\cross\B$ term also guarantees the neglect of viscous contribution in the induction equation, i.e., Eq.~(26) of \cite{PW08} guarantees Eq.~(\ref{eq:gnc}). Thus the contribution of viscous terms to the generalized Ohms law can be neglected. As a result, the induction equation remains same as in \cite{PW08}
\begin{eqnarray}
\delt \B = \curl\left[
\left(\v\cross\B\right) - \frac{4\,\pi\,\eta_O}{c}\,\J - \frac{4\,\pi\,\eta_H}{c}\,\J\cross\bb
\right. \nonumber\\
\left.
+ \frac{4\,\pi\eta_A}{c}\,
\left(\J\cross\bb\right)\cross\bb
\right]\,,
\label{eq:ind}
\end{eqnarray}
The Ohm ($\eta_O$), ambipolar ($\eta_A$) and
Hall ($\eta_H$) diffusivities are 
\bq
\eta_H = \left(\frac{v_A^2}{\omH}\right)\,,\quad
\eta_A = D\,\beta_i\,\eta_H\,,\quad
\eta_O = \beta_e^{-1}\,\eta_H\,,
\label{eq:ddf}
\eq
Eqs.~(\ref{eq:cont}), (\ref{eq:meq}) and (\ref{eq:ind}) together with the barotropic relation $P = c_s^2\,\rho$ and Amp\'ere's law 
\begin{equation}
    \J = \frac{c}{4\pi}\curl\B\,.
    \label{eq:Amp}
\end{equation}
provides the single fluid MHD description with the FLR correction of a partially ionized plasma. As can be seen from Eq.~(\ref{eq:ind}), when ions are unmagnetized, i.e., 
Hall $\beta_i<1$, Hall dominates ambipolar diffusion. The perpendicular viscosity, which is almost independent of the Hall $\beta_i$, is also important in this regime. When $\beta_i>1$, ambipolar diffusion dominates Hall. As a result, perpendicular and gyroviscous momentum transport operates together with the ambipolar diffusion in this regime. 

To summarize, the single fluid MHD description of a partially ionized plasma \citep{PW06, PW08} with the non--ideal, FLR Hall correction  has been generalized here to include the parallel, perpendicular and gyro viscosity. The parallel and perpendicular viscosity (which is mainly due to neutrals), is almost independent of the ion magnetization, i.e., the value of ion--Hall $\beta_i$. The ion magnetization determines whether the Hall diffusion or, gyroviscosity is important.  

\section{Waves in the solar atmosphere}
As has been noted in the introduction, the magnetic structures in the solar atmosphere are highly dynamic and accompanied by numerous flows and shocks of various spatial and temporal scales. Here we shall assume that the spatial scale of the flow is much smaller than the typical diameter of a flux tube. Thus for the small scale (compared with the typical diameter of a flux tube) swirls in the chromosphere, we shall approximate part of the cylindrical tube by a planar sheet and work in  Cartesian coordinates where $x\,,y\,,z$  correspond to the local  radial, azimuthal and vertical directions.  We assume an initial homogeneous state with azimuthal linear shear flow profile $\v = s \,x\,\bmath{y}$ with the shear $s={v_0}^{\prime}$ being constant. We shall also assume a uniform background field that only have a vertical component, i.e. $\B = (0, 0, B_z)$. In the presence of a purely vertical magnetic field (along $z$ in the Cartesian geometry), the dispersion relation for the waves propagating along the magnetic field is (Appendix B)
\bq
X^2+Y^2= s\,\big[\left(\om4-\omW\right)\,X + \left(\omP-\oma2\right)\,Y\big]\,.
\label{eq:drs1}
\eq
Here
\begin{eqnarray}
X&=&\sigma^2+\left(\omP+\oma2\right)\,\sigma+\omP\,\oma2-\omW\,\om4 + \omA^2\,,
\nonumber\\
Y&=&\left(\omW+\om4\right)\,\sigma+\omP\,\om4+\omW\,\oma2\,,
\end{eqnarray}
and 
\bq
\omA = k\,\va\,,
\eq
is the \alf frequency.

We can see from Eq.~(\ref{eq:drs1}) that only in the presence of Hall  ($\omW\neq 0$), or gyroviscosity ($\om4\neq 0)$, shear can channel free energy to the waves. 
Similarly, the presence of perpendicular viscosity, or ambipolar diffusion causes the damping of waves. To uncover the nature of these waves, we shall first analyze the dispersion relation, Eq.~(\ref{eq:drs1}) in the absence of shear. Thus after setting, $s=0$, Eq.~(\ref{eq:drs1}) becomes
\bq
\sigma^2+k^2\left(\nu+\eta\right)\,\sigma+k^4\,\nu\,\eta+\omA^2=0\,,
\label{eq:drss}
\eq
where
\begin{eqnarray}
\nu&=&\nu_2\pm i\,\nu_4\,,
\nonumber
\\
\eta&=&\eta_P\pm i\,\eta_H\,.
\end{eqnarray}

{\bf{Case I: $\omW=\om4=0$}} (No Hall, No gyroviscosity).\\
From  Eq.~(\ref{eq:drss}) we get the following damping roots for $\sigma=-i\,\omega$ 
\bq
\omega=\frac{-i}{2}\,\left(\oma2+\omP\right) \pm \omA\,\bigg[1-\left(\frac{\oma2-\omP}{2\omA}\right)^2\bigg]^{1/2}\
\label{eq:dmp}
\eq
Thus for small (with respect to $\omA$) $\oma2\equiv k^2\,\nu_2$ and $\omP\equiv k^2\,\eta_P$, the above dispersion relation describe an \alf wave propagting at $\omega \approx \omA$ and experiencing damping at a rate $(\oma2+\omP)/2$. For large $\oma2$ and $\omP$ there is no oscillatory solution, only damping. 

In the absence of $\eta_P$, Eq.~(\ref{eq:dmp}) provides the following cutoff wavelength due to perpendicular viscosity
\bq{}
\lambda_{c\,,\nu_2}=\frac{\pi\nu_2}{v_A}\,,
\label{eq:ct1}
\eq
below which wave can not propagate in the medium. In the photosphere-chromosphere, where $\rho\sim \rho_n$,and $\nu_2\sim \lambda_{\mbox{mfp}}\,c_s$, the cutoff wavelength becomes $\lambda_c\sim \sqrt{\beta}\,\lambda_{\mbox{mfp}}$.\\

As we see from Eq.~(\ref{eq:dmp}), ambipolar diffusion provides another cutoff wavelength 
\bq
\lambda_{c\,,\eta_A}=\pi\,\frac{\va}{\nu_{ni}}\,,
\label{eq:amc}
\eq
Thus only waves above cut--off wavelength
\bq
\lambda_c=max(\lambda_{c\,,\nu_2}\,,\lambda_{c\,,\eta_A})\,,
\eq
can propagate in the photosphere-chromosphere without damping.

\begin{figure}
\includegraphics[scale=0.35]{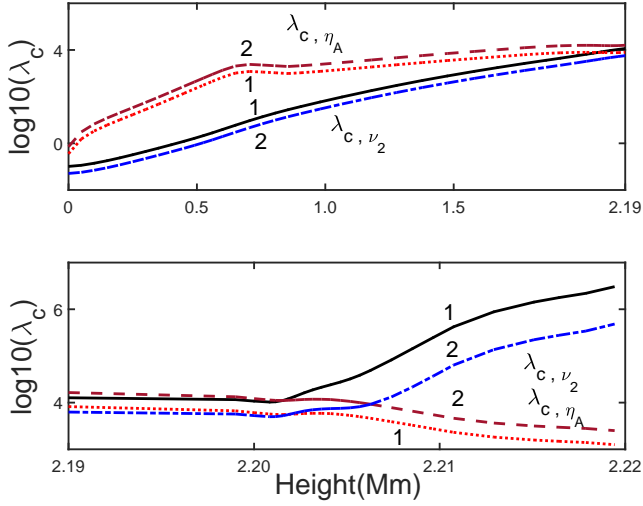}
\caption{The cut--off wavelength $\lambda_C$ due to perpendicular viscosity, $\lambda_{c\,,\nu_2}$  and  due to ambipolar diffusion, $\lambda_{c\,,\eta_A}$  against height is plotted for $B_0=50\,\mbox{G}$ (curves labelled $1$) and $B_0=100\,\mbox{G}$ (curves labelled $2$). The altitude dependence of the magnetic field, density and temperature is the same as in the previous figures.}
 \label{fig:FN4}  
\end{figure}
In Fig.~(\ref{fig:FN4}) we plot the cut--off wavelength $\lambda_c$ due to perpendicular viscosity, $\lambda_{c\,,\nu_2}$ and ambipolar diffusion, $\lambda_{c\,,\eta_A}$ against height for $B_0=50\,\mbox{G}$ (curves labelled $1$) and $B_0=100\,\mbox{G}$ (curves labelled $2$). We notice that for 
$B_0=100\,\mbox{G}$ field damping in the chromosphere is mainly due to ambipolar diffusion  while for a weaker ($50\,\mbox{G}$) field, ambipolar and perpendicular viscosity in equal measure damps the wave in the upper chromosphere. In the transition region it is  perpendicular viscosity that is responsible 
for wave damping. To summarize, for weaker field, both ambipolar and perpendicular viscosity are important to the wave damping in the upper chromosphere while  perpendicular viscosity is the dominant damping mechanism in the transition region. For stronger field, perpendicular viscosity is important only in the transition region. Therefore, the role of perpendicular FLR viscosity (which is mainly  due to neutrals)  may become important to the wave heating in this region. 

\begin{figure}
\includegraphics[scale=0.35]{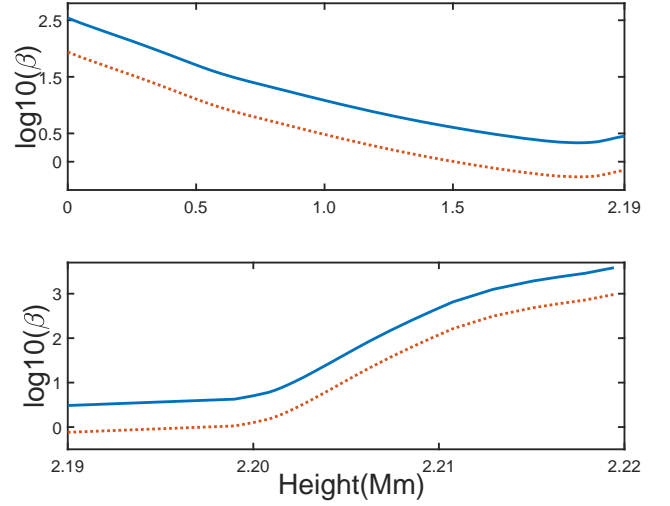}
\caption{The plasma $\beta=2\,c_s^2/\va^2$  is plotted against the height for $B_0=100\,\mbox{G}$ (solid curve) and $B_0=200\,\mbox{G}$ (dotted curve)  field. The other parameters are same as in previous figures.}
 \label{fig:FN5}  
\end{figure}

{\bf{Case II: $\om4\neq 0$}} (When gyroviscosity is present).\\
In the absence of Pedersen diffusion and perpendicular viscosity, the dispersion relation Eq.~(\ref{eq:drss}) describe polarized Hall and whistler waves. We analyse Eq.~(\ref{eq:drss}) in the following limiting cases before subjecting  it to numerical solution.   

In the absence of gyroviscosity, i.e., when $\nu_4=0$ we recover the low ($\omega\ll\omA$)  and high ($\omA\ll \omega$) frequency left circularly polarized ion--cyclotron ($\omega=\omH$) and  right circularly polarized whistler ($\omega=\omW=k^2\,\eta_H$) waves from Eq.~(\ref{eq:drss}).

When $\nu_4\neq 0$ and $\eta_H=0$, in the low frequency ($\omega\ll\omA$) limit, when the magnetic and gyroviscous stress balance each other, Eq.~ (\ref{eq:drss}) gives
\bq
\omega_{gh}=\frac{\va^2}{\nu_4}\simeq\frac{G(\rho)}{\beta}\,\omH\,,
\label{eq:lfw}
\eq
which is scaled, left circularly polarized Hall wave. We call this low frequency branch  gyro-Hall wave, $\omgh$ to distinguish it from the Hall wave, $\omega_H$.  
Here
\bq
G(\rho)=8\,\left(\frac{\rho}{\rho_i}\right)^2\,,
\eq
is proportional to the density ratio and is $\sim \mathcal{O}(1)$ in the upper chromosphere and transition region. 

The high frequency ($\omA\ll \omega$) frequency branch (when the magnetic stress is unimportant and fluid inertia balances the gyroviscous stress) becomes
\bq
\omega_{gw}=\omega_{4}\simeq\frac{\beta} {G(\rho)}\,\omW\,,
\label{eq:hfw}
\eq
which describes the scaled right circularly polarized whistler ($\omW=k^2\,\eta_H$) wave. We call this branch  gyro-whistler wave, $\omgw$.

In the above equations we have used $\nu_4=\eta_{i\,3}/2\simeq (\beta/8) (\rho_i/\rho)^2\,\eta_H$ valid in the $\beta_i\gg 1$ limit.  
As $\rho\sim \rho_i$ in the transition region, two branches of gyroviscous wave are scaled (by plasma $\beta$) Hall branches. With increasing magnetic field strength, the plasma $\beta$ decreases (Fig.~\ref{fig:FN5}). However, even for very strong field at the footpoint, plasma $\beta>1$ in the transition region. Thus the gyroviscous wave propagates at frequencies much higher than the whistler frequency, $\omW$. To summarize, both Hall and gyroviscous effect excite similar waves in the solar atmosphere except that they operate in different ion-Hall window and at different frequencies. 

To see that the gyro-Hall and gyro-whistler waves are indeed left and right circularly polarized waves, from Eq.~(\ref{eq:fdm}) and Eq.~(\ref{eq:fdm1}), we get
\bq
 \left(
\begin{array}{cc} X & Y\\
                    
                 -Y   & X    
  \end{array}
\right)\,\frac{\delta \v_{\perp}}{\va}
= 0\,.
\label{eq:cp}
\eq
From the dispersion relation, Eq.~(\ref{eq:drs1}), we have $X=\pm i\,Y$. Plugging it back in the Eq.~(\ref{eq:cp}) leads to
\bq
\delta v_y=\pm i\,\delta v_x\,,
\eq
i.e., $\delta v_y$ leads (lags) $\delta v_x$ by $\pi/2$. Clearly velocity fluctuations, in general are the superposition of left-circularly polarized gyro-Hall, $\delta v_{x1}$ and right-circularly polarized gyro-whistler $\delta v_{x2}$ waves
\bq
\delta v = \delta v_{x1}\left(1\,,i\,,0\right)\,e^{i\,\left(\omega_{gh}\,t+k\,z\right)}+\delta v_{x2}\left(1\,,-i\,,0\right)\,e^{i\,\left(\omega_{gw}\,t+k\,z\right)}\,.
\eq

\begin{figure}
\includegraphics[width=9cm, height=6cm]{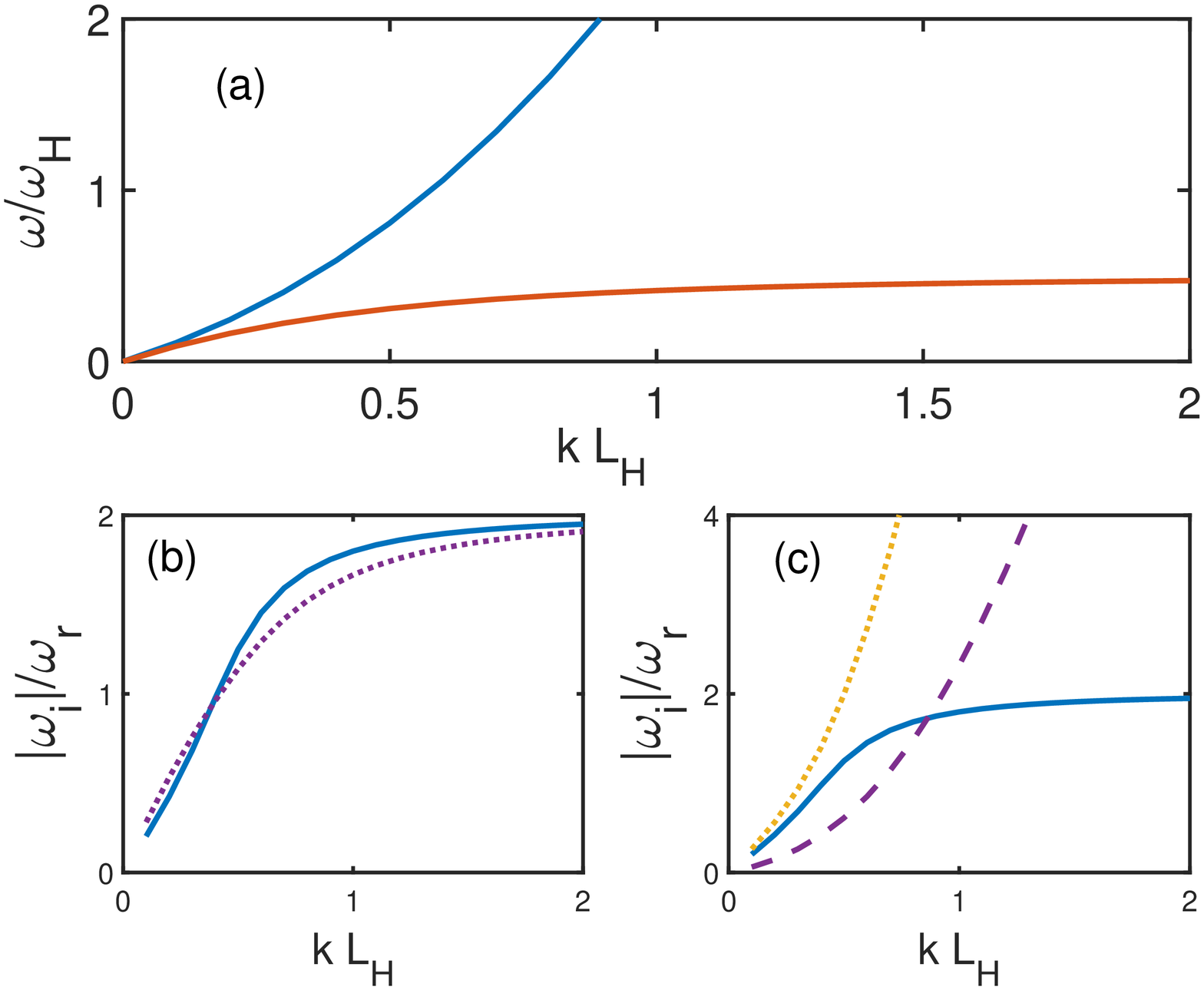}
\caption{In panel (a) $\omega/\omega_H$ is plotted against $k\,L_H$ by setting $\eta_H=\eta_P=\nu_2=0$ and $\beta=G(\rho)$
and in panels (b) and (c) for $\nu_2=2\,\nu_4$, ratio of the absolute value of $Im(\omega)=\omega_i$ to $Re(\omega)=\omega_r$ is plotted for the gyro-whistler and gyro-Hall waves respectively.}
 \label{fig:FNW}  
\end{figure}

Numerical solution of the Eq.~(\ref{eq:drss}) is shown in Fig.~(\ref{fig:FNW}). As expected, the gyroviscosity, like Hall lifts  \alf degeneracy and waves are splits into low-frequency gyro-Hall (for negative sign before $\nu_4$ term in the dispersion relation (\ref{eq:drss})) and high-frequency gyro-whistler waves (for positive sign before $\nu_4$ term)  [panel (a) in the figure]. However, perpendicular viscosity $\nu_2=2\,\nu_4$ causes severe damping of the gyro-whistler wave [solid curve, panel (b)]. When ambipolar diffusion is switched on, i.e. take $\beta_i=2$, damping rate is not significantly impacted (dotted curve). Therefore, the gyro-whistler wave is primarily damped by the perpendicular viscosity. On the other hand, ambipolar diffusion is the main damping agent of the gyro-Hall wave [panel (c)]. The solid curve in the panel (c) is for $\nu_2=2\,\nu_4$ (but without ambipolar), dashed curve is for $\nu_2=0\,,\beta_i=1$ and dotted curve is for $\nu_2=2\,\nu_4\,,\beta_i=1$.            
By setting $\nu_2=\eta_H=\eta_A=0$ but in the presence of shear. dispersion relation Eq.~(\ref{eq:drs1}) becomes
\bq
\sigma^4+ C_2\,\sigma^2+C_0=0\,,
\label{eq:nW2}
\eq
where
\bq
C_2=2\,\omA^2-s\,\omega_4+\omega_4^2\,,\quad C_0=\omA^4(1-s\frac{\nu_4}{\va^2})\,.
\eq
Note that the gyroviscosity channels the shear energy to waves. Like Hall, the presence of shear flow destabilizes gyroviscous wave if $C_0<0$, 
or
\bq
s>\omgh\,\,.
\label{eq:gIC}
\eq
Instability is caused by the gyroviscous momentum transport across the magnetic field due to finite Hall frequency and is similar to the ion-cyclotron instability of \cite{Q15}. 
Except for the numerical factor ($\sim \beta$), the onset condition for the gyroviscous, Eq.~(\ref{eq:gIC}) and Hall, Eq,~(\ref{eq:hins}) are exactly opposite of each other. However, like Hall, gyroviscous instability depends on the orientation of the magnetic field. Therefore, for a given orientation of the magnetic field in the network and internetwork region the sign of flow gradient will determine whether Hall,  or gyroviscosity will  channel the shear energy to the waves. 

Why is the onset condition  for the gyroviscous and Hall instabilities, Eq.~(\ref{eq:hins}) and Eq.~(\ref{eq:gIC}) have opposite sign? This is due to the fact that while 
shear generates $\delta b_y$ from $\delta b_x$, it acts as a sink for $\delta v_y$ due to  $\delta v_x$. This is reflected in the sign difference of the shear term in the momentum Eq.~(\ref{eq:fdm}) and induction Eq.~(\ref{eq:fdm1}).   

Adopting $\va\,,\nu_4$ as units, we may write $\omA=k\,,\omega_4=k^2$ and the dispersion relation, Eq.~(\ref{eq:nW2}) becomes
\bq
\sigma^4+(k^2+2-s)k^2\,\sigma^2+k^4(1-s)=0\,.
\label{eq:nW2a}
\eq
The discriminant of the above equation is
\bq
\Delta=k^4\big[\left(k^2-s\right)^2+4\,k^2\big]\equiv k^4\,r^2\,.
\label{eq:dist}
\eq
From the growth rate of the instability   
\bq
\sigma=\left(\frac{2\left(s-1\right)\,k^2}{k^2+r-(s-2)}\right)^{1/2}\,,
\label{eq:gr}
\eq
we derive the following limiting cases
\bq
\sigma =
\begin{cases} 
      k\,\sqrt{s-1}\,, & \text{for small $k$, $r\approx s$}, \\
      \sqrt{s-1}\,, & \text{for $k\gg 1$, $r\approx k^2$}\,.
\label{eq:asmt}
\end{cases}
\eq
Here $\sigma$ and $s$ are in the unit of $\va^2/\nu_4$. Thus in dimensional form, the above equation is
\bq
\sigma =
\begin{cases} 
      k\,\sqrt{s\,\nu_4-\va^2}\,, & \text{for small $k$, $r\approx s$}, \\
      \left(\frac{\va^2}{\nu_4}\right)^{1/2}\sqrt{s-\frac{\va^2}{\nu_4}}\,, & \text{for $k\gg 1$, $r\approx \omega_4$}\,.
\label{eq:asmt1}
\end{cases}
\eq

\begin{figure}
\includegraphics[width=9cm, height=6cm]{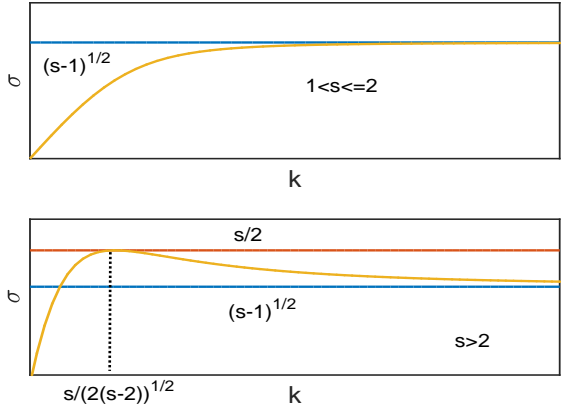}
\caption{The growth rate $\sigma$ against $k$ is plotted for $1<s\leq 2$ (upper frame) and $s>2$ (lower frame).}
 \label{fig:FN6}  
\end{figure}
In Fig.~(\ref{fig:FN6}, top frame) growth rate, $\sigma(k)$ is plotted for $1<s\leq 2$. For small $k$ the growth rate, in conformity with Eq.~(\ref{eq:asmt}), grows linearly with $k$. However, the growth rate saturates quickly and asymptotically approaches constant $\sqrt{s-1}$ for large $k$. As there is no dissipation in the present case, the growth rate remains constant at all wavelengths. 

\begin{figure}
\includegraphics[width=9cm, height=4cm]{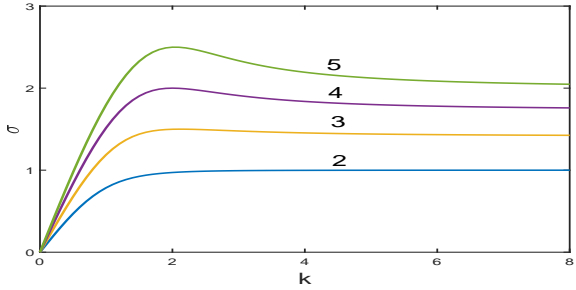}
\caption{The growth rate $\sigma$ against $k$ is plotted for $s=2\,,3\,,4\,,5$ .}
 \label{fig:FNN8}  
\end{figure}
The growth rate has a  maximum only if $s>2$. In order to see this we rewrite the dispersion relation, Eq.~(\ref{eq:nW2a}) as $a\,k^4+b\,k^2+c=0$ with 
\bq
a=1-s+\sigma^2\,,b=\left(2-s\right)\sigma^2\,,c=\sigma^4
\label{eq:akf}
\eq
and set the discriminant $b^2-4\,a\,c=0$ which gives the maximum growth rate
\bq
\sigma_0=\frac{s}{2}\,.
\label{eq:mgr}
\eq
Except for the sign, this is identical to the maximum Hall instability growth rate \citep{PW12}. See  Eq.~(\ref{eq:Hgr}) below. The wavenumber, corresponding to the maximum growth rate is
\bq
k_0=\left(\frac{s/2}{\left(s-2\right)}\right)^{1/2}\,.
\label{eq:kgr}
\eq

In Fig.~(\ref{fig:FN6}, lower frame) the growth rate. $\sigma(k)$ is plotted for $s>2$. We see that $\sigma(k)$ increases linearly with $k$, reaches maximum at $k_0$ before asymptotically approaching $\sqrt{s-1}$.

In Fig.~(\ref{fig:FNN8}) we plot the growth rate, Eq.~(\ref{eq:gr}) against $k$ for different shear rate $s$ which is labelled against the curve. The growth rate increases with increasing $k$, reaching maximum at $k_0$ before declining asymptotically to $\sqrt{s-1}$ at large $k$. With increasing shear energy, i.e., with increasing $s$, the growth rate increases.        

Like Hall, Eq.~(\ref{eq:nW2}) can be analysed in  different limits. 

(I.) {\bf Cyclotron limit:} Momentum transport due to gyroviscosity is balanced by the fluid advection.  This is the short wavelength ($\omA\gg \sigma$) limit in which we set $\sigma=0$ in Eq.~(\ref{eq:fdm}) (Appendix B), or $a=0$ in Eq.~(\ref{eq:akf}). The instability condition is 
same as Eq.~(\ref{eq:gIC}), The instability growth rate is $\sqrt{s-1}$ (Eq.~\ref{eq:asmt}).

(II) {\bf Gyrovisocus limit:} Momentum transport due to gyroviscosity in this case is balanced by the fluid inertia. This is the long wavelength ($k\ll1$) limit in which from Eq.~(\ref{eq:fdm}) (Appendix B) we get $\sigma^2+(2-s)\,k^2=0$, or
\bq
\sigma=k\,\sqrt{s-2}\,,
\eq
i.e., the long wavelength fluctuations grows only if $s>2$. Plugging $k=k_0$ from Eq.~(\ref{eq:kgr}) in the above equation, we recover maximum growth rate, $s/2$ (Eq.~\ref{eq:mgr}).

{\bf{Case II: $\omega_2\neq 0\,,\omega_P\neq 0\,,\omW=0$}}\\
\begin{figure}
\includegraphics[scale=0.35]{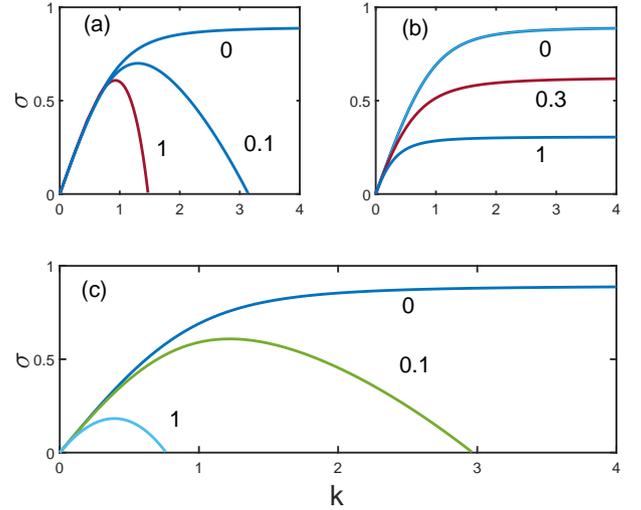}
\caption{The growth rate $\sigma$ against $k$ for $s=1.8$ is plotted for (a) $\nu_2/\nu_4=0\,,0.1\,,1$, and $\eta_P/\nu_4=0$, (b) $\nu_2/\nu_4=0$, and $\eta_P/\nu_4=0\,,0.3\,,1$, (c) $\nu_2/\nu_4=\eta_P/\nu_4=0\,,0.1\,,1$.} 
\label{fig:FNN9}  
\end{figure}
For comparison in Fig.~\ref{fig:FNN9} we plot the growth rate of the gyroviscous instability when  $\omega_2=\omega_P=0$ (curve labeled  by $0$). In the presence of perpendicular viscosity ($\omega_2=k^2\,\nu_2\neq 0$) but without magnetic diffusion ($\omega_P=0$),  gyroviscous instability grows only at long wavelengths [Fig.~\ref{fig:FNN9}(a)] with the cutoff determined by
\bq
k_c^2=\frac{1}{\nu_2}\left(\frac{2\left(s-1\right)\,k^2}{k^2+r-(s-2)}\right)^{1/2}\,.
\label{eq:gr}
\eq
Thus we see that with increasing $\nu_2/\nu_4$ from $0.1$ to $1$ gyroviscous instability stops growing at shorter wavelengths.   

In the presence of magnetic diffusion ($\omega_P\neq 0$) but without perpendicular viscosity ($\omega_2=0$) [Fig.~\ref{fig:FNN9}(b)], the instability keeps growing at all wavelengths with the asymptotic behaviour given by Eq.~(\ref{eq:asmt}). However, as Ohm and ambipolar diffusion are dissipative in nature, the growth rate $\sigma$ decreases with increasing $\omega_P$.  

When $\omega_2=\omega_P\neq 0$ [Fig.~\ref{fig:FNN9}(c)], instability grows at a much lower rate than when perpendicular viscosity and Pedersen diffusion operates separately. The cutoff wavenumber is shifted to slightly lower values compared to [Fig.~\ref{fig:FNN9}(a)]. However, the change to wavenumber is not significant compared to the growth rate.        

{\bf{Case III: Hall case ($\omega_4=\omega_2=\omega_P=0$)}}\\
Although the effect of Ohm, Hall and ambipolar diffusion on the linear wave propagation in the solar atmosphere has been investigated in the past \citep{PW08, P08, PW12, PW13}, we briefly revisit this topic here. 

The dispersion relation, Eq.~(\ref{eq:drs1})  reduces to Eq.~(\ref{eq:nW2}) except now   
\bq
C_2=2\,\omA^2+s\,\omW+\omW^2\,,\quad C_0=\omA^4\,\left(1+ \frac{s}{\omH}\right)\,.
\label{eq:Hc}
\eq
In the absence of shear flow, i.e. $s=0$ we recover the low ($\omega\ll\omA$)  and high ($\omA\ll \omega$) frequency left circularly polarized ion--cyclotron ($\omega=\omH$) and  right circularly polarized whistler ($\omega=\omW=k^2\,\eta_H$) waves respectively.The presence of shear flow destabilises these waves only if $C_0<0$, or
\bq
-s>\omH\,,
\label{eq:hins}
\eq
with the growth rate   
\bq
\sigma=\left(\frac{-2\left(s+\omH\right)\,\omA^2}{s+2\,\omH+\omW+r}\right)^{1/2}\,.
\label{eq:grH}
\eq
Here $r^2=(\omW+s)^2+4\,\omA^2$. From Eq.~(\ref{eq:grH}) we see that for $\omA/\omH\ll 1$, $r\approx s$ and $\sigma=-i\,\omA$, i.e., purely oscillatory mode. As it is only Hall that can channel the free shear energy to the waves, it is not surprising that in the $\omA/\omH\ll 1$ limit, which correspond to the first quadrant of Fig.~1 in \cite{P18}, there is no growth rate but pure \alf oscillation. 

When $\omA/\omH\gg 1$, $r\approx \omW$ and  from Eq.~(\ref{eq:grH}) we obtain
\bq
\sigma \approx \left(-s\,\omH\right)^{1/2}\,,
\eq
which we recognise as the growth rate of the ion-cyclotron wave [Eq.~(53) in \cite{PW13}].

The maximum growth rate 
\bq
\sigma_0=\frac{-s}{2}\,,
\label{eq:Hgr}
\eq
is reached at 
\bq
k_0=\left(\frac{-s^2/2\,\eta_H}{s+2\,\omH}\right)^{1/2}\,.
\eq

In the {\bf cyclotron ($\sigma\ll\omA$)} limit,  the dispersion relation, Eq.~(\ref{eq:nW2a}) with coefficients, Eq.~(\ref{eq:Hc}) becomes
\bq
(2\,\omA^2+s\,\omW+\omW^2)\,\sigma^2 + \omA^4\,\left(1+ \frac{s}{\omH}\right)=0\,,
\label{eq:nW2c}
\eq
with the instability criteria, Eq.~(\ref{eq:hins}). 

In the {\bf diffusive ($\sigma\gg\omA$)} limit, the dispersion relation, Eq.~(\ref{eq:nW2a}) with coefficients, Eq.~(\ref{eq:Hc}) becomes
\bq
\sigma^2+\omW\,(s+\omW)=0\,,
\label{eq:nW2W}
\eq
with the necessary condition for the instability becoming
\bq
-s>\omW\equiv k^2\,\eta_H\,
\label{eq:hins1}
\eq
while the maximum growth rate is given by Eq.~(\ref{eq:Hgr}).

The effect of perpendicular viscosity, $\omega_2$ and ambipolar diffusion is same as in the gyroviscous case. 

\section{discussion}
The identification and understanding of the wave propagation and heating in the partially ionized solar atmosphere is a major problem in the solar physics. Quantitative MHD modelling of the weakly ionized and weakly magnetized photosphere and weakly ionized and moderately (strongly) magnetized middle (upper) chromosphere suggests that the non-ideal MHD effect such as Ohm, Hall and ambipolar diffusion are important to the wave propagation and heating in the chromosphere, transition region and corona.  

The one dimensional modelling of the solar atmosphere suggests that in the photosphere and lower and middle chromosphere, the neutral Hydrogen density far exceeds the plasma number density whereas in the upper chromosphere and  transition region the plasma number density exceeds neutral hydrogen density by orders of magnitude \citep{F93}. This fact coupled with the changing ion magnetization changes the scale and nature of plasma transport in the solar atmosphere.  For example, the ion Larmor radius becomes a function of the fractional ionization and resulting finite Larmor radius effect manifests over macroscopic ($\sim$ few to few hundred kilometer) scales. Thus not only the induction equation but momentum equation should also reflect this non-ideal feature in the partially ionized photospehere-chromosphere and transition region.  

Non-ideal magnetic diffusion (via Ohm, ambipolar and Hall) and viscous momentum transport (due to parallel, perpendicular and gyroviscous terms in the pressure tensor) competes with each other throughout the solar atmosphere. The parallel and perpendicular viscous momentum transport which is  caused mainly by the neutrals and is almost independent of the ion magnetization, competes with the Hall and ambipolar diffusion of the magnetic field in the photosphere and lower and middle chromosphere. In fact for a weak ($<50\,\mbox{G}$) magnetic field at the footpoint,  the perpendicular viscosity may dominate ambipolar diffusion in the upper chromosphere.         

The gyroviscosity is like Hall, except it is caused by the magnetized ions. Thus the upper chromosphere is subject to two, comparable non--ideal MHD effect: one causing the ambipolar slippage of the ions against the sea of neutrals (resulting in magnetic diffusion) and the other causing the gyroviscous transport of the momentum. In the transition region however, ambipolar diffusion disappears, and owing to the strong ion magnetization ($\beta_i\gg1$) another non--dissipative, non--ideal MHD effect, namely gyroviscous transport of the momentum becomes important. The magnetic flux through the surface area $\vec{S}$ of the tube $\int \vec{B}\cdot d\vec{S}$ is frozen at this altitude. 

In the solar quite regions of the upper chromosphere and transition layers ($B_0\sim 100\,\mbox{G}$), where the ions are strongly magnetized ($\beta_i\gg1$), both the left and right circularly polarized gyroviscous waves are excited in the medium at frequencies much smaller (larger) than the corresponding Hall (whistler) frequencies. Since plasma beta, $\beta \gtrsim 10^3$ in this region, left circularly polarized  gyroviscous waves [Eq.~(\ref{eq:lfw})] will propagate at a much lower (about thousand times) frequency than the right polarized [Eq.~(\ref{eq:hfw})] waves.  As $\eta_H=10^8\,\mbox{cm}^2\,s^{-1}$ for $B_0=100\,G$ (Fig.~\ref{fig:FN7D}) we see from Eq.~(\ref{eq:hfw}) that $\omega \sim 10^{12}\,k^2$.  In the active regions, for a kG field, the frequency can increase by couple of order of magnitude. Thus it should be possible the detect few hundred km wavelength waves of frequencies $\omega\sim 10^{3}-10^{4}\,\mbox{s}^{-1}$ in the upper chromosphere and transition region.        

Solar prominences and filaments are relatively cool ($\sim 10^4\,\mbox{K}$) partially ionized large--scale  magnetic structures located in the corona. More like cosmic abundances they are composed of atomic hydrogen ($\sim 90\%$) and Helium. The fractional ionization in the prominences is unknown and could be as low as $0.1$ or, as large as $10$ \citep{PV02}. As the ions are magnetized it would appear that not only the ambipolar diffusion \citep{S09} but the gyroviscous effect will also affect the wave propagation in the filaments. Assuming $B=5\,\mbox{G}$ and $\rho=5\times10^{-14}\,\mbox{g}\,\mbox{cm}^{-3}$ we get $\va\sim 10-100\,\mbox{km}/\mbox{s}$. For typical $c_s\sim 10\,\mbox{km}\,{s}^{-1}$ the plasma $\beta \lesssim 1$. Thus both the low and high  frequency gyroviscous waves may propagate in the prominences and filaments.

Although direct signature of the gyroviscous waves may be difficult to detect, these waves may become unstable in the presence of vortex motion in the fluid. Both the Hall and gyroviscous instabilities require favourable velocity gradient. As $\beta$ is large in the transition region [Fig.~{\ref{fig:FN5})], any favourable field geometry will ensure that the onset condition of the instability, $s\gtrsim\omH/\beta$ [Eq.~(\ref{eq:gIC})] is easily met. Note that the Hall diffusion of the magnetic field and gyroviscous momentum transport operate in different ion magnetization windows and the onset condition of the Hall and gyroviscous instabilities are mutually exclusive.

As the presence of swirls of various sizes have been observed in the chromosphere and transition region the gyroviscous effect may excite low frequency turbulence. Observations and numerical simulations suggest ubiquitous presence of flow gradients in the photospheric--chromospheric plasma \citep{Y21, B08, At09, W09, Z93, SN98}. The typical vorticity of a vortex is $\sim 6\times 10^{-3}\,\persec$ corresponding to a rotation period of $\sim 35\,$ minutes \citep{B10}. Thus it would appear that the gyroviscous instability does not have time to develop since the growth rate ($|{v_0}^{\prime}| / 2 = 3\times 10^{-3}\,\persec$) is very small. However, above vorticity value is limited by the upper limit in the vorticity resolution ($\sim 4\times 10^{-2}\,\persec$, \cite{B10}). The numerical simulation gives much higher vorticity value ($\sim 0.1-0.2\,\persec$) in the photosphere--lower chromosphere (Fig.~31, \cite{SN98}). The growth rate corresponding to $|{v_0}^{\prime}| = 0.2\,\persec$ is one minute. Therefore, it is quite likely that the gyroviscous waves in the transition region will become unstable in the presence of shear flow given that the maximum wavelength corresponding to the maximum growth rate fits within the pressure scale height ($\sim 150\,\mbox{km}$), i.e. $k_0\,h\gtrsim 1$.  

Note that in this work we have assumed linear shear profile. However, recent numerical simulation \citep{Y21}, suggest that a cubic profile is better suited to the rotational flows. 
Closer to the photosphere, the linear shear profile is off by about $10\,\%$ from the simulated vortex profiles. It is only at the base ($\sim 0.5\,\mbox{Mm}$) of the chromosphere that the error becomes $\sim 5\,\%$. As gyroviscous instability is important in the transition ($\gtrsim 2.0  \mbox{Mm}$) region, we conclude that the linear shear profile is a good approximation to the vortex flows.  

The net radiative loss from the chromosphere $\sim 10^7\,\mbox{erg}\,\mbox{cm}^{-2}\,\mbox{s}^{-1}$ is an order of magnitude greater than that of the overlying corona. To heat the chromosphere to $10^4\,\mbox{K}$ ten times more energy flux is required than to heat the corona to million degree. It is believed that most of the solar heating takes place in the chromosphere where the convection energy is transported by some physical mechanism. It is quite plausible that the present gyroviscous shear instability may contribute towards heating by driving the turbulence in the chromosphere. However, only numerical simulations can provide a definitive answer to what role the gyroviscous shear instability will have in the heating of the chromosphere. 

\section{summary}
The partially ionized solar photosphere--chromosphere mainly consists of neutral Hydrogen with a modest fraction of Helium. The ratio of plasma to  neutral number density is very low in the medium. This tiny fraction of charged particle undergo frequent collision and also as frequently they exchange their identity with the neutrals. Thus the magnetic stress is easily passed to the neutrals over collision/charge exchange time scale which results in various non--ideal MHD effects \citep{PW08}. Not only does the magnetic field diffuse through the partially ionized medium due to the non--ideal MHD effects but owing to collision and charge exchange, the non--diagonal pressure tensor terms in the momentum equation also become important. This leads on the one hand to the viscous damping of the waves in the photosphere-chromosphere, and on the other hand to the propagation of the polarized gyroviscous waves in the transition region. In the presence of shear flow gradient, these polarized waves may become unstable in the upper chromosphere and the transition region. The presence of favourable magnetic field topology will facilitate the transfer of shear energy to the magnetic fluctuations.          

The following are the itemized summary of the present work.

1. The effective Larmor radius in a partially ionized plasma is a function of fractional ionization. Thus the finite Larmor radius effect, which manifests as viscous momentum transport, operates over macroscopic scale in the solar atmosphere. 

2. The parallel and perpendicular viscosity is important in the photosphere-chromosphere and competes with the non-ideal Pedersen and Hall magnetic diffusion. In the upper chromosphere perpendicular viscosity may dominate ambipolar diffusion while in the transition region where ambipolar diffusion disappears, dominant viscosity is gyroviscosity.    

3. In the presence of a purely vertical magnetic field  polarized gyroviscous wave may propagate undamped along the field line in the transition region. 

4. In the presence of shear flow gradient these  waves may become unstable. The maximum growth rate of the instability ($\sim$ few minute) is proportional to the shear gradient. 

5. Although the maximum growth rate of the gyroviscous instability is identical to the Hall instability \citep{PW12} the onset condition in the two cases are opposite of each other.

6. As the gyroviscous waves of few hundred km wavelength falls within the current observational resolution ($\sim 100\, \mbox{km}$), it should be possible to identify these waves and associated instabilities in the transition region. \\

\section*{Acknowledgements}
This research was supported by the Australian Government through the Australian Research Council's Discovery Projects funding scheme (project DP190103100).

\section*{Data availability}
There are no new data associated with this article.

\appendix
\section[]{The viscous tensor components in a partially ionized fluid}

The detailed derivation of the ion and neutral viscosity coefficients, $\eta$ are given by \cite{Z02}. Assuming $T_i = T_e= T_n = T$, the ion and neutral viscosity coefficients are 
\begin{eqnarray}
\eta_{i\,0} = \frac{p_i}{2\,\nu_i}\,\xi_i\,\Delta_{\eta}^{-1}\,,\quad
\eta_{i\,1} = \frac{\eta_{i\,0}}{1+\Delta_{\eta}^{-2}\,\beta_i^2}\,,
\nonumber\\  
\eta_{i\,2} = \eta_{i\,1}\Big[\frac{\beta_i}{2}\Big]\,,
\quad
\eta_{i\,3} = \eta_{i\,1}\,\beta_i\,\Delta_{\eta}^{-1}\,,
\label{eq:ivis}
\end{eqnarray}
and
\begin{eqnarray}
\eta_{n\,0} = \frac{p_n}{2\,\nu_n}\,\xi_n\,\Delta_{\eta}^{-1}\,,\quad
\eta_{n\,1} =\eta_{n\,0}\, \frac{1+\beta_i^2\,\xi_n^{-1}\,\Delta_{\eta}^{-1}
}{1+\beta_i^2\,\Delta_{\eta}^{-2}}\,,
\nonumber\\  
\eta_{n\,2} = \eta_{n\,1}\Big[\frac{\beta_i}{2}\Big]\,,
\quad
\eta_{n\,3}=\eta_{n\,0}\,\,\beta_i\,\Delta_{\eta}^{-1}\,\frac{1-\left(\frac{\Delta_{\eta}}{\xi_n}\right)}{1+\left(\frac{\beta_i^2}{
\Delta_{\eta}^2}\right)}\,,
\label{eq:nvis}
\end{eqnarray}
\bq
\eta_{i\,,n\,4} = \eta_{i\,,n\,3}\Big[\frac{\beta_i}{2}\Big]
\eq
where the square bracket $[\beta_i/2]$ means that wherever $\beta_i$ occurs they should be replaced  by $\beta_i/2$ to get the new viscosity coefficients. Here we use slightly more general definition of Hall beta than in Eq.~(\ref{eq:IhB}), i.e., $\beta_i=\omega_{ci}/\nu_i$. 

The collision frequencies in the above expressions are
\bq
\nu_{\alpha} = 0.3\,\nu_{\alpha\,\alpha} + \sum_{\substack{\beta\neq \alpha}} f_{\alpha\,\beta}\,\nu_{\alpha\,\beta}\,,
\eq
where
\bq
f_{\alpha\,\beta} = \frac{m_{\alpha}\,m_{\beta}}
{\left(m_{\alpha}+m_{\beta}\right)^2}\,\Big[1+0.6\left(\frac{m_\beta}{m_\alpha}\right)\,A_{\alpha\,\beta}^{*}
\Big]\,.
\eq
Defining
\bq
g_{\alpha\,\beta}=\frac{m_{\alpha}\,m_{\beta}}
{\left(m_{\alpha}+m_{\beta}\right)^2}\,\Big[1-0.6\left(\frac{m_\beta}{m_\alpha}\right)\,A_{\alpha\,\beta}^{*}
\Big]\,,
\eq
we may write other parameters in the viscosity coefficients as 
\begin{eqnarray}
\xi_{\alpha}&=& 1+g_{\alpha\,\beta}\,\left(\frac{\nu_{\alpha\,\beta}}{\nu_{\beta}}\right)\,,
\nonumber\\
\Delta_{\eta}&=&1-g_{\alpha\,\beta}\,g_{\beta\,\alpha}\frac{\nu_{\alpha\,\beta}\nu_{\beta\,\alpha}}{\nu_{\alpha}\nu_{\beta}}
\end{eqnarray}

For hard sphere collision model $A_{\alpha\,\beta}^{*} =1$. As solar photosphere is a mixture of atomic hydrogen ($90\%$) and Helium ($10\%$), the helium mass density is $0.1\times 4$ hydrogen density, $\rho_H$. Thus the total gas density $\rho=1.4\,\rho_H$. The gas number density, $n$ in terms of hydrogen number density $n_H$ is $n=1.1\,n_H$. This results in the average neutral mass $m_n=\rho/n=1.3\,m_H$. Assuming $m_i=m_p$ and $m_n=1.3\,m_p$ where $m_p=1.67\times10^{-24}\,\mbox{g}$ is the proton mass, we get $f_{i\,n}=0.44\,,f_{n\,i}=0.36$ and $f_{i\,e}=0.5\times 10^{-3}\,,f_{n\,e}=0.4\times 10^{-3}$. Thus the collision frequencies becomes
\begin{eqnarray}
\nu_i &=& 0.3\,\nu_{ii} + 0.44\,\nu_{in} + + 0.5\times 10^{-3}\,\nu_{ie}\,,
\nonumber\\
\nu_n &=& 0.3\,\nu_{nn} + 0.36\,\nu_{ni} + 0.4\times 10^{-3}\,\nu_{ne}\,.
\label{eq:cfx}
\end{eqnarray}

The electron-ion collision frequency $\nu_{ei}$ is \citep{Z02}
\bq
\nu_{ei}=\frac{8\,\sqrt{\pi}}{3} \frac{e^4\,n_e\,L}{m_e^{1/2}}\,
\left(k_B\,T\right)^{-3/2}\,\mbox{s}^{-1}\,,
\eq
where Coulomb logarithm, $L=ln(3\,k_B\,T\,\lambda_D/e^2)$ with 
the Debye length $\lambda_D=\sqrt{4\,\pi\,n_e\,e^2/k_B\,T}$. For $L=22$ in the photosphere-chromosphere, $\nu_{ei}$ can be
expressed in terms of fractional ionization $X_e$  
\bq
 \nu_{ei}=98\,X_e\,n_n\,T^{-3/2}\,\mbox{s}^{-1}\,   
\eq
where $\mbox{T}$ and $n_n$ are in $\mbox{K}$ and $\mbox{cm}^{-3}$ respectively. Note that the collision frequency, $\nu_{ei}$ of \cite{Z02} is twice the value of \citep{BR65} and in \cite{PW08} owing to slightly different definition of the collision relaxation time.    

Thus
\bq
\nu_{ie} = \left(\frac{m_e}{m_i}\right)\,\nu_{ei} \equiv 5.35\times 10^{-2}\,X_e\,n_n\,T^{-3/2}\,\mbox{s}^{-1}\,.
\eq
The ion--ion collision frequency is 
\bq
\nu_{ii}= \sqrt{\frac{2\,m_e}{m_i}}\,\nu_{ei}=3.24\,X_e\,n_n\,T^{-3/2}\,\mbox{s}^{-1}\,,
%0.5\,\sqrt{\frac{m_e}{m_i}}\,\nu_{ei}\equiv 1.17\times %10^{-2}\,\nu_{ei}\,.
\label{eq:nuii}
\eq
where we have assumed $n_i=n_e$.

The neutral--neutral collision frequency is 
\bq
\nu_{nn} = n_n\,\sigma\,v_{th}\,\mbox{s}^{-1}\,,
\eq 
where $v_{th}$ is the thermal speed of the neutral particle. Taking hydrogen cross-section $\sigma\approx 10^{-14}\,\mbox{cm}^{-2}$ \citep{V16}, neutral-neutral rate coefficient for $m_n=1.3\,m_p$ becomes
\bq
\sigma\,v_{th}=7.97\times 10^{-11}\,T^{\frac{1}{2}}\quad\mbox{cm}^3\,\mbox{s}^{-1}\,.
\eq
 
The plasma--neutral collision frequency $\nu_{j n}$ is \citep{WK93}
\begin{equation}%%%%12
\nu_{j n} = \gamma_{j n}\,\rho_n  = \frac{\sigv_j}{m_n + m_j}\,\rho_n \,.
\end{equation}
Here $\sigv_j$ is the rate coefficient for the momentum transfer by
collision of the $j^{\mbox{th}}$ particle with the neutrals.  The  $H^+-H$ collision 
cross-section is energy dependent [Fig.~12, \cite{PG08}] and for $[0.5-10]\,eV$ which is relevant to the photosphere--transition region $\sigma \approx 10^{-14}\,\mbox{cm}^2$. The $e-H$ collision 
cross-section, Fig.~14 \citep{PG08} for low ($1 \lesssim eV$) energies is $\sigma \approx 4\times 10^{-15}\,\mbox{cm}^2$ whereas for $[1-10]\,eV$, $\sigma \approx 10^{-15}\,\mbox{cm}^2$. We adopt 
$\sigma \approx 2.12\times 10^{-15}\,\mbox{cm}^2$ which reproduces \cite{D83} electron-neutral rate coefficient. Thus the ion-neutral and electron-neutral rate coefficients are 
\begin{eqnarray}
<\sigma\,v>_{in} &=& 0.5 \cross 10^{-10}\,T^{\frac{1}{2}}\quad\mbox{cm}^3\,\mbox{s}^{-1}\nonumber \\
<\sigma\,v>_{en} &=& 8.28 \cross 10^{-10}\,T^{\frac{1}{2}}\quad \mbox{cm}^3\,\mbox{s}^{-1}\,.
\end{eqnarray}
The corresponding collision frequencies are 
\begin{eqnarray}
\,\nu_{in} = 0.51\times 10^{-10}\,n_n\,\,T^{\frac{1}{2}}\,\mbox{s}^{-1}\,,\nonumber\\
\,\nu_{en} = 8.28\times 10^{-10}\,n_n\,\,T^{\frac{1}{2}}\,\mbox{s}^{-1}
\,. 
\label{eq:cf2} 
\end{eqnarray} 
Taking account of the various plasma--neutral collision frequencies we may write Eq.~(\ref{eq:cfx}) as
\begin{eqnarray}
\nu_i &\approx& \left(0.44+ 1.9\times 10^{10}\,X_{e}\,T^{-2}\right)\,\nu_{in}\,. 
\label{eq:cfx1}
\end{eqnarray}
Below $1\,\mbox{Mm}$ in the solar atmosphere, $\nu_i \approx \nu_{in}$. Between $1-2.1\,\mbox{Mm}$, 
$\nu_{ii}>\nu_{in}$. As the neutral density plummets in the transition region, $\nu_{in}$ becomes smaller by orders of magnitude compared to $\nu_{ii}$. 

The ratios of ion and neutral viscosity coefficients, Eq.~(\ref{eq:ivis})-(\ref{eq:nvis}) are 
\begin{eqnarray}
\frac{\eta_{i\,1}}{\eta_{i\,0}} \sim \frac{1}{\beta_i^2}, \quad \frac{\eta_{i\,3}}{\eta_{i\,0}} \sim  \frac{1}{\beta_i}
\nonumber\\
\frac{\eta_{n\,1}}{\eta_{n\,0}} \sim \mathcal{O}(1), \,\, \frac{\eta_{n\,3}}{\eta_{n\,0}} \sim \frac{\xi_n-\Delta_{\eta}}{\beta_i}\,.
\end{eqnarray}
As the perpendicular $-\bf{b}\cross(\bf{b}\cross \nabla$) and gyroviscous, $\bf{b}\cross \nabla$ ion viscosity coefficients $\eta_{i\,1}$, and $\eta_{i\,3}$ are $1/\beta_i^2$ and $1/\beta_i$ smaller than the parallel coefficient, $\eta_{i\,0}$, the ion gyroviscosity dominates the ion perpendicular viscosity when  $\beta_i>1$.  The neutral perpendicular viscosity coefficient, $\eta_{n\,1}$ is comparable to the parallel coefficient, $\eta_{n\,0}$ while neutral gyroviscous term $\eta_{n\,3}$ is negligible compared to the $\eta_{n\,0}$.    
\begin{figure}
\includegraphics[scale=0.35]{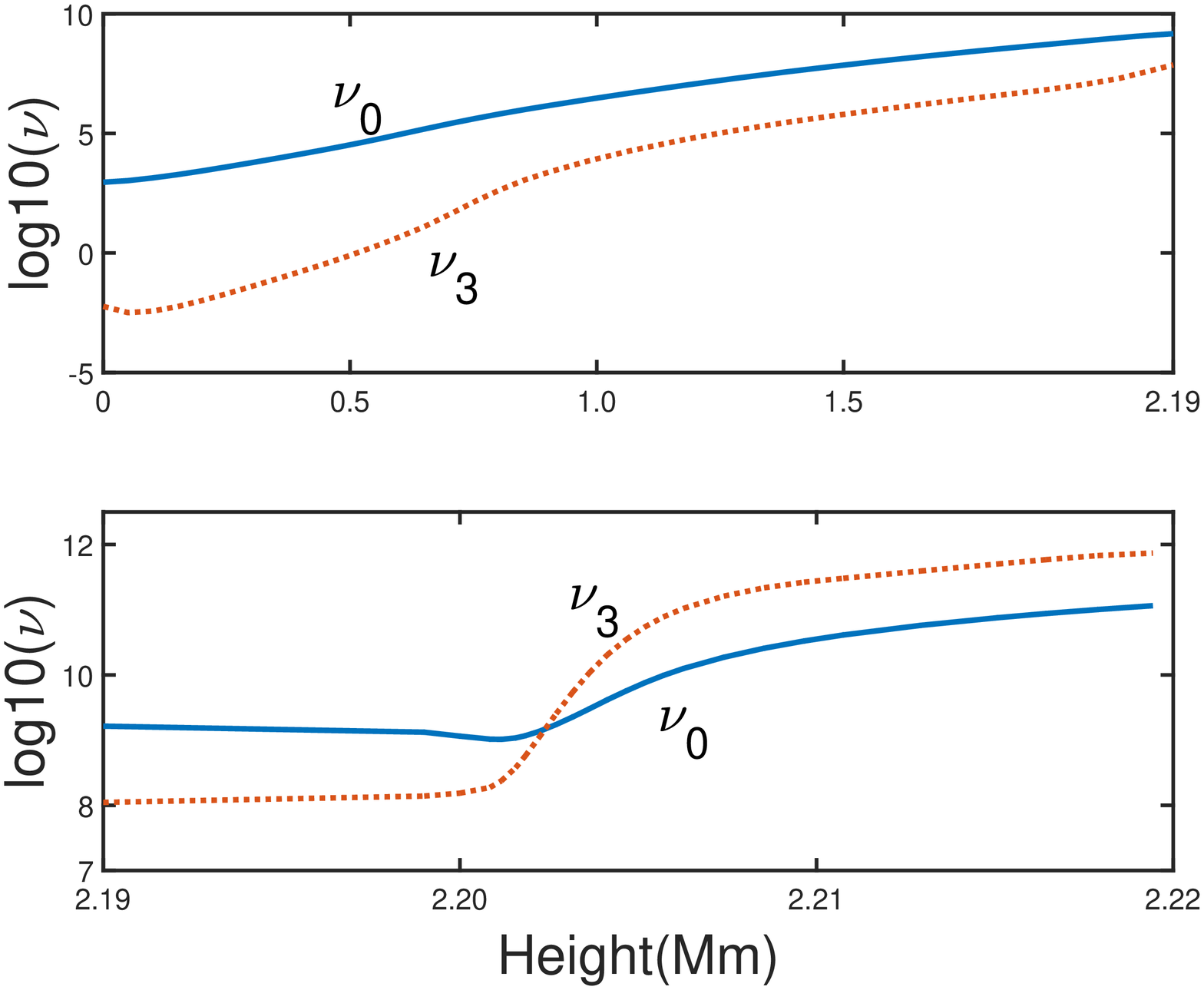}
\caption{The parallel, $\nu_{0}$ (solid line), and gyroviscosity  $\nu_{3}$ (dotted line) are plotted against the height.  Other parameters are the same as in the previous figures.} 
\label{fig:FV}  
\end{figure}
The ratio of ion and neutral parallel viscosity coefficients is 
\bq
\frac{\eta_{i\,0}}{\eta_{n\,0}}\sim 
\frac{n_i}{n_n}\frac{\nu_n}{\nu_i}\,. 
\eq
In the photosphere-chromosphere, where neutral hydrogen density is the dominant component of the partially ionized 
gas, $\eta_{n\,0}$ is orders of magnitude larger than  $\eta_{i\,0}$ (Fig.~\ref{fig:Fpar}). However, in the transition region, where neutral number density sharply drops, $\eta_{i\,0}$ becomes orders of magnitude larger than $\eta_{n\,0}$ [Fig.~\ref{fig:Fpar} (lower frame)]. As the parallel viscosity, $\nu_{0}$ is the  sum of $\eta_{i\,0}$ and $\eta_{n\,0}$, it keeps steadily growing and dominates gyrovisocity in the 
photosphere-chromosphere Fig.~\ref{fig:FV}. The perpendicular viscosity, $\nu_1$ and $\nu_2$ are identical to $\nu_0$ owing to $\eta_{n\,1\,,2}\sim \eta_{n\,0}$. Therefore, both the parallel ($\nu_0$) and perpendicular ($\nu_1\,,\nu_2$) vicosity dominats gyroviscosity in the  photosphere-chromosphere. 

The ratio of ion and neutral gyroviscosity coefficient $\eta_3$ is
\bq
\frac{\eta_{i\,3}}{\eta_{n\,3}} \sim \left(\frac{\eta_{i\,0}}{\eta_{n\,0}}\right)\frac{1}{\xi_n-\Delta_{\eta}}\,.
\eq
As $\xi_n-\Delta_{\eta}$ is close to zero in the photosphere-chromosphere and is $0.1$ in the transition region, $\eta_{i\,3}$  dominants $\eta_{n\,3}$ throughout the solar atmosphere. Thus, the gyroviscosity, $\nu_3$ is mainly due to the ion gyroviscosity and dominates $\nu_0\,,\nu_1$ and $\nu_2$ in the transition region (Fig.~\ref{fig:FV}, lower frame). To summarize, the parallel and perpendicular viscosity is the dominant FLR viscosity in the photosphere-chromosphere whereas cross FLR viscosity is dominant FLR effect in the transition region. 

Various components of the strain $W_{\alpha\,\beta}$ are 
\begin{eqnarray}
\tenq{W}_{0}&=&  \frac{3}{2}\,\left(\bb\cdot\tenq{W}\cdot\bb\right)\left(\bb\bb-\frac{1}{3}\tenq{\bf{I}}\right)\,,
\nonumber\\
\tenq{W}_{1}&=&\left(\tenq{\bf{I}}-\bb\bb\right)\cdot\tenq{W}\cdot\left(\tenq{\bf{I}}-\bb\bb\right)
-\frac{1}{2}\left(\tenq{\bf{I}}-\bb\bb\right) \left(\tenq{\bf{I}}-\bb\bb\right)\bmath{:}\tenq{W}\,,
\nonumber\\
\tenq{W}_{2}&=&\left(\tenq{\bf{I}}-\bb\bb\right)\cdot\tenq{W}\cdot \bb\bb + \mbox{Transpose}\,,
\nonumber\\
\tenq{W}_{3}&=&\frac{1}{2}\Big[\bb\cross\tenq{W}\cdot\left(\tenq{\bf{I}}-\bb\bb\right) + \mbox{Transpose}\Big]\,,
\nonumber\\
\tenq{W}_{4}&=& \bb\cross\tenq{W}\cdot\bb\bb + \mbox{Transpose}\,.
\label{eq:wab}
\end{eqnarray}

In the component form $\tenq{W}$ is
\begin{eqnarray}
W_{xx}&=&\frac{4}{3}\partial_x\vx-\frac{2}{3}\left(\partial_y\vy+\partial_z\vz\right)\,, 
\nonumber\\
W_{yy}&=&\frac{4}{3}\partial_y\vy-\frac{2}{3}\left(\partial_x\vx+\partial_z\vz\right)\,,
\nonumber\\
W_{zz}&=&\frac{4}{3}\partial_z\vz-\frac{2}{3}\left(\partial_x\vx+\partial_y\vy\right)\,, 
\end{eqnarray}
and the remaining symmetric part is 
\begin{eqnarray}
W_{xy} &=& \partial_x\vy+\partial_y\vx\,, 
\nonumber\\
W_{xz} &=& \partial_x\vz+\partial_z\vx\,, 
\nonumber\\
W_{yz}&=& \partial_y\vz+\partial_z\vy\,. 
\end{eqnarray}

In the presence of a vertical magnetic field, i.e. $\bb=\bm{z}$, 
\bq
\bb\bb-\frac{1}{3}\tenq{\bf{I}}=
\left(
\begin{array}{ccc} -1/3 &  0 & 0\\
                    0 & -1/3 & 0\\
                      0    &  0     & 2/3    
  \end{array}
\right)\,.
\label{eq:fd1}
\eq
As $\bb\cdot\tenq{W}\cdot\bb=W_{zz}$ we have
\bq
\tenq{W}_{0}=\left(
\begin{array}{ccc} -W_{zz}/2 &  0 & 0\\
                    0 & -W_{zz}/2 & 0\\
                      0    &  0     & W_{zz}    
  \end{array}
\right)\,.
\label{eq:tW0}
\eq
Similarly 
\bq
\left(\tenq{\bf{I}}-\bb\bb\right)\cdot\tenq{W}\cdot\left(\tenq{\bf{I}}-\bb\bb\right)=
\left(
\begin{array}{ccc} W_{xx} &  W_{xy} & 0\\
                    W_{yx} & W_{yy} & 0\\
                      0    &  0     & 0    
  \end{array}
\right)\,.
\eq
As $\left(\tenq{\bf{I}}-\bb\bb\right)\bmath{:}\tenq{W}=W_{xx}+W_{yy}$ we have
\bq
\left(\tenq{\bf{I}}-\bb\bb\right) \left(\tenq{\bf{I}}-\bb\bb\right)\bmath{:}\tenq{W}
=\left(
\begin{array}{ccc} W_{xx}+W_{yy} &  0 & 0\\
                    0 & W_{xx}+W_{yy} & 0\\
                      0    &  0     & 0    
  \end{array}
\right)\,.
\eq
Thus
\begin{eqnarray}
 \tenq{W}_{1}&=&
\left(
\begin{array}{ccc} \frac{1}{2}\left(W_{xx}-W_{yy}\right) & W_{xy} & 0\\
                    W_{yx} & -\frac{1}{2}\left(W_{xx}-W_{yy}\right) & 0\\
                      0    &  0     & 0    
  \end{array}
\right)\,,
\nonumber
\\
 \tenq{W}_{2}&=&
\left(
\begin{array}{ccc} 0 & 0 & W_{xz}\\
                    0 & 0 & W_{yz}\\
                      W_{zx}    &  W_{zy}     & 0    
  \end{array}
\right)\,.
\label{eq:tW2}
\end{eqnarray}
As
\bq
 \bb\cross\tenq{W} = \left(
\begin{array}{ccc} -W_{yx} &  -W_{yy} & -W_{yz}\\
                    W_{xx} & W_{xy} & W_{xz}\\
                      0    &  0     & 0    
  \end{array}
\right)\,,
\label{eq:fd1}
\eq
we have
\bq
 \bb\cross\tenq{W}\cdot\left(\tenq{\bf{I}}-\bb\bb\right) = \left(
\begin{array}{ccc} -W_{yx} &  -W_{yy} & 0\\
                    W_{xx} & W_{xy} & 0\\
                      0    &  0     & 0    
  \end{array}
\right)\,,
\label{eq:fd1}
\eq
Adding the above matrix with its transpose gives
\bq
  \tenq{W}_{3}=\frac{1}{2}\left(
\begin{array}{ccc} -2\,W_{xy} &  W_{xx}-W_{yy} & 0\\
                    W_{xx}-W_{yy} & 2\,W_{xy} & 0\\
                      0   &  0     & 0    
  \end{array}
\right)\,,
\label{eq:tW3}
\eq
For $\tenq{W}_{4}$ note that
\begin{eqnarray}
\bb\cross\tenq{W}\cdot\bb\bb=
\left(
\begin{array}{ccc} -W_{yx} &  -W_{yy} & -W_{yz}\\
                    W_{xx} & W_{xy} & W_{xz}\\
                      0    &  0     & 0    
  \end{array}
\right)\times
\nonumber\\
\left(
\begin{array}{ccc} 0 &  0 & 0\\
                    0 & 0 & 0\\
                      0   &  0     & 1    
  \end{array}
\right)
=\left(
\begin{array}{ccc} 0 &  0 & -W_{yz}\\
                    0 & 0 & W_{xz}\\
                      0    &  0     & 0    
  \end{array}
\right)\,.
\end{eqnarray}
Adding with its transpose gives
 \bq
 \tenq{W}_{4}=\left(
\begin{array}{ccc} 0 &  0 & -W_{yz}\\
                    0 & 0 & W_{xz}\\
                      -W_{yz}    &  W_{xz}     & 0    
  \end{array}
\right)\,.
\label{eq:tW4}
 \eq
The above Eqs.~(\ref{eq:tW0}),  (\ref{eq:tW2}), (\ref{eq:tW3}), and (\ref{eq:tW4}) are also given on page 252 by \cite{BR65}.

When the ion {\it acquires} neutral inertia due to its frequent collision and/or due to rapid charge exchange with the neutrals, ion gyro frequency and consequently ion gyro radius becomes a function of fractional ionization. Furthermore, the ion fluid moves under the combined pressure $P_i + P_n$ \citep{HW98}.  Thus 
\bq
\eta_{i\,3} \approx \frac{P_i+P_n}{2\,\omH}\,,
\eq  
and $\eta_{i\,4} \approx 2\,\eta_{i\,3}$. Defining 
\bq 
f=\frac{1+X_e}{1+2\,X_e}\,,
\eq
which varies between one and two third between the photosphere and transition region, we may write $P_i+P_n=P\,f$. Assuming $P=c_s^2\,\rho$ we may write $\eta_{i\,3}$ in terms of $\nu_{3}$ as
\bq
\nu_{3}=\frac{1}{2}\,{\rl}^2\,\omH\,.
\eq
Thus $\nu_{4}=2\,\nu_{3}$ as $\eta_{i\,4}=2\,\eta_{i\,3}$. 

Combining the last two terms in Eq.~(\ref{eq:vst}) we get 
\bq
\tenq{\Pi}_{\Lambda} = \frac{1}{2}\nu_{3}
\Big[\bb\cross\tenq{W}\cdot\left(\tenq{\bf{I}}+3\bb\bb\right) + \mbox{Transpose}\Big]\,,
\eq
which in matrix form becomes
\bq
\frac{\tenq{\Pi}_{\Lambda}}{0.5\,\nu_{3}} = \left(
\begin{array}{ccc} -2\,W_{yx} &  W_{xx}-W_{yy} & -4\,W_{yz}\\
                    W_{xx}-W_{yy} & 2\,W_{xy} & 4\,W_{xz}\\
                      -4\,W_{yz}   &  4\,W_{xz}     & 0    
  \end{array}
\right)\,.
\eq
Note that gyroviscosity has nothing to do with the viscosity as the associated viscous stresses, Eq.~(\ref{eq:stn}) are always perpendicular to the velocity, implying that there is no viscous dissipation associated with $\tenq{\Pi}_{\Lambda}$.

Lastly, the various components of the symmetric tensor $\tenq{\Pi}$ are
\begin{eqnarray}
 \Pi_{xx}/\rho&=&\frac{\nu_0}{2}\,W_{zz}-\frac{\nu_1}{2}\,\left(W_{xx}-W_{yy}\right)-\nu_3\,W_{xy}\,,
 \nonumber\\
 \Pi_{yy}/\rho&=&\frac{\nu_0}{2}\,W_{zz}+\frac{\nu_1}{2}\,\left(W_{xx}-W_{yy}\right)+\nu_3\,W_{xy}\,,
 \nonumber\\
 \Pi_{zz}/\rho&=&-\nu_0\,W_{zz}
 \nonumber\\
 \Pi_{xy}/\rho&=&-\nu_1\,W_{xy}+\frac{\nu_3}{2}\,\left(W_{xx}-W_{yy}\right)\,,
 \nonumber\\
 \Pi_{xz}/\rho&=&-\nu_2\,W_{xz}-2\nu_3\,W_{yz}\,,
 \nonumber\\
 \Pi_{yz}/\rho&=&-\nu_2\,W_{yz}+2\nu_3\,W_{xz}\,.
\end{eqnarray}
Making use of $W_{zz}=-(W_{xx}+W_{yy})$ in $\Pi_{xx}$ and $\Pi_{yy}$ the above equation reduces to equation (2.21) of \cite{BR65}.  

\section[]{Waves in a homogeneous medium}
We shall assume a partially ionized solar plasma threaded 
by the vertical magnetic field, $\bb=\bm{z}$. As the Larmor scale over which the various FLR viscosity affects the transport properties of the plasma is large in the transition region, we shall focus on the role of ion viscosity, $\nu_{i\,3}$ on the wave propagation in the medium. We provide a brief discussion on the effect of parallel viscosity towards the end of this appendix.   
We want to derive a linear dispersion relation in the presence of gyroviscous effect. To that end note that the components of $\div \Pi\,$ are
 \begin{eqnarray}
(\div \Pi)_x/\rho=\frac{\nu_0}{2}\partial_xW_{zz}- \nu_1\,\nabla_{\perp}^2\vx-\nu_2\,\partial_z W_{xz}
\nonumber\\
-\nu_{3}\,\left(\nabla_{\perp}^2\vy+2\,\partial_z W_{yz}\right)\,, 
\nonumber\\
(\div \Pi)_y/\rho=\frac{\nu_0}{2}\partial_yW_{zz}- \nu_1\,\nabla_{\perp}^2\vy-\nu_2\,\partial_z W_{yz}
\nonumber\\ 
+ \nu_{3}\,\left(\nabla_{\perp}^2\vx+2\,\partial_z W_{xz}\right)\,,
\end{eqnarray}
and
\begin{eqnarray}
(\div \Pi)_{z}/\rho =-\nu_0\,\partial_z W_{zz}-\nu_2\,\Big[\nabla_{\perp}^2\vz
+\partial_z \left(\partial_x\vx+\partial_y\vy \right)\Big]
\nonumber\\ 
-2\,\nu_{3}\,\partial_z\left(\partial_x\vy-\partial_y\vx\right)\,.
\end{eqnarray}
Here $\nabla_{\perp}^2=\partial_x^2+\partial_y^2$. After linearizing Eqs.~(\ref{eq:meq}) and (\ref{eq:ind}) and assuming that $\delta\,f \propto \exp{\left(\sigma\,t + i\,k\,z\right)}$ for the transverse components we have
%\begin{eqnarray}
\bq
 \left(
\begin{array}{cc} \sigma+\omega_2 & \omega_4\\
                    
                 s-\omega_4   & \sigma+\omega_2    
  \end{array}
\right)\,\frac{\delta \v_{\perp}}{\va}
%\nonumber\\
= i\,\omA\,\delta {\bb}_{\perp}
\,,
\label{eq:fdm}
\eq
%\end{eqnarray}
and
\bq
\left(
\begin{array}{cc} \sigma+\omega_P & \omega_W\\
                    
                  -\left(s+\omega_W\right)   &    \sigma+\omega_P  \end{array}
\right)\,\delta {\bb}_{\perp}
= i\,\omA\,\frac{\delta \v_{\perp}}{\va}
\,.
\label{eq:fdm1}
\eq
Here $\eta_P=\eta_O+\eta_A$ is the Pedersen diffusion coefficient, $\omega_{2}=k^2\,\nu_{n\,2}\,,\omega_{4}=k^2\,\nu_{i\,4}$, and the whistler and Pedersen frequencies are, $\omega_W=k^2\,\eta_H\,,\omega_P=k^2\,\eta_P$. From Eqs.~(\ref{eq:fdm}) and (\ref{eq:fdm1}) we get the dispersion relation, Eq.~(\ref{eq:drs1}).

The longitudinal component of the dispersion relation becomes
\bq
\omega^2-k^2\,c_s^2-\frac{4}{3} i\,\omega\,k^2\,\nu_0=0\,,
\eq 
which describes the damped sound waves
\bq
Re(\omega)^2=k^2\,c_s^2-\frac{4}{9}\left(k^2\,\nu_0\right)^2
\eq
with the damping rate
\bq
Im(\omega)=\frac{2}{3} k^2\,\nu_0\,.
\eq

\section[]{Magnetic diffusivities}

\begin{figure}
\includegraphics[scale=0.35]{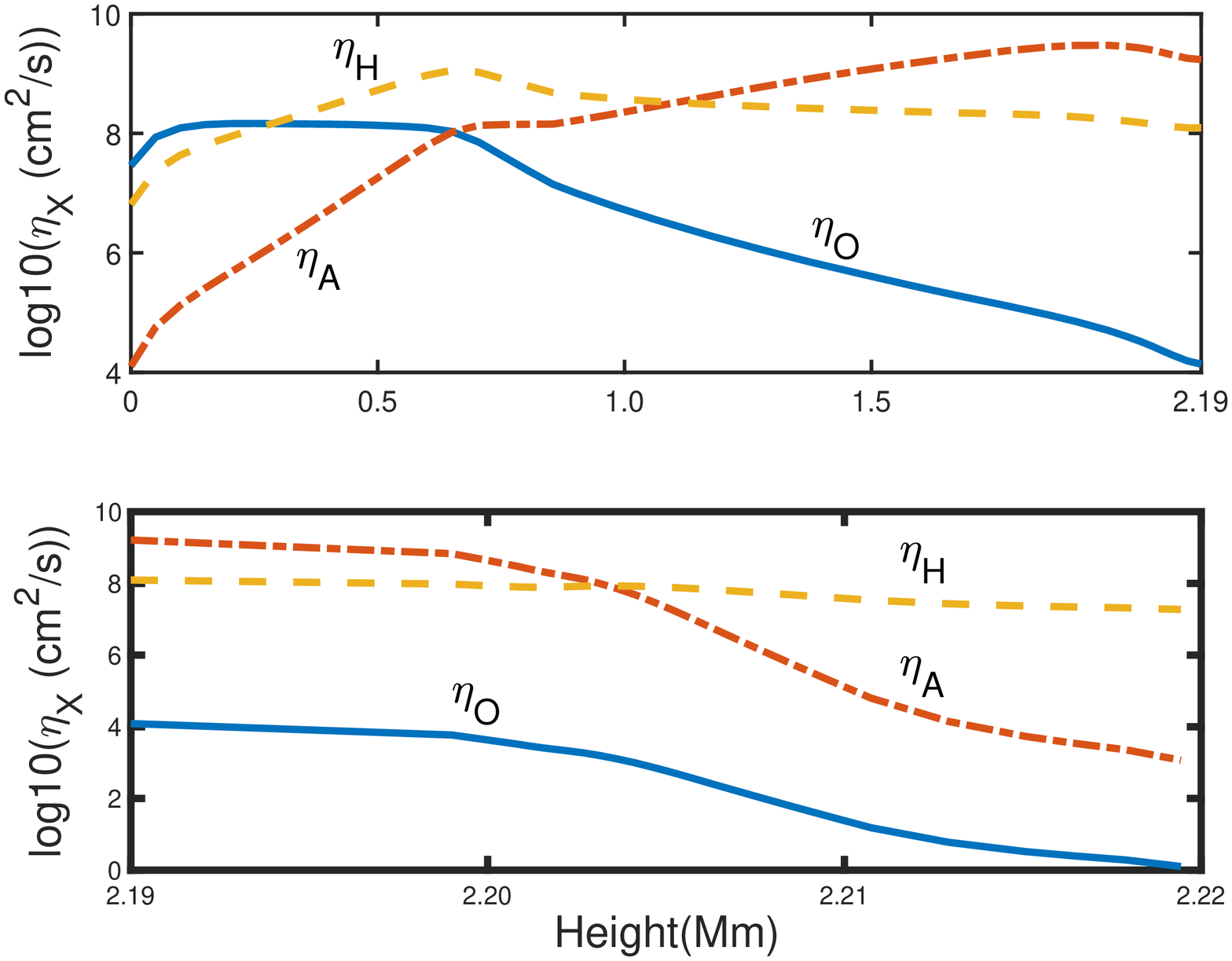}
\caption{The Ohm ($\eta_O$), Hall ($\eta_H$) and ambipolar ($\eta_A$) diffusion are plotted against the height for $B_0=100\,\mbox{G}$ and $m_i=m_p$, $m_n=1.3\,m_p$. Hall scales with the magnetic field 
as $\propto B$ while ambipolar scales as $B^2$.  Other parameters are the same as in the previous figures.} 
\label{fig:FN7D}  
\end{figure}
In Fig.~(\ref{fig:FN7D}), we plot magnetic diffusivities against height for $B_0=100\,G$.
Ohm diffusion, $\eta_O$ is independent of the magnetic field whereas Hall, and ambipolar scales as $\eta_H\propto B\,,\eta_A\propto B^2$ with the magnetic field. Thus in the active regions of the sun, i.e. in the presence of stronger ($\sim kG$) field, ambipolar is more likely to dominate the Hall diffusion whereas in the quiet region of the sun Hall may dominate ambipolar.   

Note that the diffusivities are also given in \cite{PW12, PW13}, where  we have used $m_i=30\,m_p$, $m_n=2.3\,m_p$ which is relevant to the molecular clouds. For atomic hydrogen ($90\%$)-Helium($10\%$) mixture in the solar photosphere, $m_i=m_p$, $m_n=1.3\,m_p$ which has been used in the present case. Further, the modified collision rates \citep{PG08} are also used. The data of Fig.~(\ref{fig:FN7D}) is given in table~\ref{table:T1}.

\begin{center}
\begin{table}
\centering
\begin{tabular}{|c c c c|}
 \hline
 height&$\logten\eta_O/\mbox{cm}^2\,s^{-1}$& $\logten\eta_H/\mbox{cm}^2\,s^{-1}$&$\logten\eta_A/\mbox{cm}^2\,s^{-1}$\\[0.5ex]
\hline
 0&7.4620 &6.8113 &4.1025 \\ 
 50&7.9352 &7.3746&4.7562  \\ 
 100&8.0910&7.6353&5.1218 \\
 150&8.1465&7.8051&5.4058  \\
 200&8.1633&7.9428&5.6645 \\
 250&8.1649&8.0703&5.9179 \\
 300&8.1621&8.1984&6.1768  \\
350&8.1576&8.3272&6.4389\\
400&8.1537&8.4600&6.7086\\
450&8.1468&8.5934 &6.9821\\
490&8.1371&8.6980&7.2010\\
525&8.1261&8.7885&7.3931\\
560&8.1138&8.8782&7.5848\\
600&8.0922&8.9733&7.7966\\
650&8.0313&9.0535&8.0179\\
705&7.8525&9.0225&8.1346\\
755&7.6121&8.9083&8.1466\\
805&7.3750&8.7928&8.1523\\
855&7.1475&8.6806&8.1553\\
905&6.9903&8.6348&8.2203\\
980&6.7752&8.5812&8.3275\\
1065&6.5499&8.5334&8.4558\\
1180&6.2709&8.4828&8.6305\\
1278&6.0529&8.4499&8.7782\\
1378&5.8423&8.4180 &8.9179\\
1475&5.6546&8.3936 &9.0470\\
1580&5.4546&8.3647 &9.1726\\
1670&5.2924&8.3439&9.2732\\
1775&5.1125&8.3252&9.3833\\
1860&4.9587&8.3009 &9.4487\\
1915&4.8490&8.2768 &9.4724\\
1980&4.7011&8.2373 &9.4741\\
2017&4.6008&8.2065 &9.4543\\
2043&4.5231&8.1823 &9.4301\\
2062&4.4606&8.1624 &9.4045\\
2075&4.4143&8.1481 &9.3820\\
2087&4.3699&8.1351 &9.3590\\
2110&4.2878&8.1124 &9.3122\\
2140&4.1914&8.0926 &9.2561\\
2168&4.1374&8.0921&9.2350\\
2190&4.0895&8.0936 &9.2095\\
2199&3.7707&7.9793 &8.8291\\
2200&3.6319&7.9344 &8.6417\\
2200.84&3.5144&7.9022&8.4753\\
2201.16&3.4649&7.8942&8.4026\\
2201.56&3.4101&7.8930&8.3203\\
2201.83&3.3776&7.8979&8.2712\\
2202.22&3.3313&7.9078&8.2006\\
2202.70&3.2704&7.9211&8.1086\\
2203.16&3.1948&7.9276&7.9926\\
2203.63&3.1036&7.9277&7.8518\\
2204.13&2.9944&7.9225&7.6848\\
2204.66&2.8622&7.9058&7.4754\\
2205.18&2.7244&7.8854&7.2596\\
2205.70&2.5783&7.8590&7.0280\\
2206.27&2.4120&7.8263&6.7661\\
2207.40&2.0923&7.7548&6.2535\\
2208.53&1.7782&7.6805&5.7510\\
2209.65&1.4751&7.6055&5.2657\\
2210.75&1.1828&7.5311&4.7978\\
2212.92&0.7771&7.4305&4.1477\\
2215.08&0.5171&7.3692&3.7308\\
2216.43&0.3983&7.3430&3.5408\\
2217.88&0.2854&7.3178&3.3598\\ 
2219.43&0.1027&7.2718&3.0679\\[1ex]
  \hline
  \label{table:T1}
\end{tabular}
\caption{The Ohm ($\eta_O$), Hall ($\eta_H$) and ambipolar ($\eta_A$) diffusion profiles for $B_0=100\,G$ field. Other parameters are same as in the figures.}
\end{table}
\end{center}

\end{document}